\documentclass[aps,prl,twocolumn,amsmath,amssymbs,superscriptaddress]{revtex4-2}

\usepackage{graphicx}
\usepackage{amsmath,amsfonts,amssymb}
\usepackage{epsfig}
\usepackage{wrapfig}
\usepackage{bbm}
\usepackage[usenames]{color}
\usepackage{array}
\usepackage{times}
\usepackage{float}

\def\kex{\kappa_{ex}}
\def\kin{\kappa_{in}}

\def\w{\omega}
\def\wo{\omega_{0}}

\def\wp{\omega_{p}}

\def\OR{\Omega_{\text{R}}}
\def\OSR{\Omega_{\text{SR}}}

\def\TR{\tau_{\text{R}}}
\def\TSR{\tau_{\text{SR}}}

\def\E{\mathcal{E}}

\def\a{a}
\def\ad{a^{\dag}}

\begin{document}

\title{Topological Frequency Combs and Nested Temporal Solitons}
\author{Sunil Mittal}
\email[Email: ]{mittals@umd.edu}
\affiliation{Joint Quantum Institute, NIST/University of Maryland, College Park, MD 20742, USA}
\affiliation{Department of Electrical and Computer Engineering, and IREAP, University of Maryland, College Park, MD 20742, USA}

\author{Gregory Moille}
\affiliation{Microsystems and Nanotechnology Division, Physical Measurement Laboratory, National Institute of Standards and Technology, Gaithersburg, MD 20899, USA}
\affiliation{Joint Quantum Institute, NIST/University of Maryland, College Park, MD 20742, USA}

\author{Kartik Srinivasan}
\affiliation{Microsystems and Nanotechnology Division, Physical Measurement Laboratory, National Institute of Standards and Technology, Gaithersburg, MD 20899, USA}
\affiliation{Joint Quantum Institute, NIST/University of Maryland, College Park, MD 20742, USA}

\author{Yanne K. Chembo}
\affiliation{Department of Electrical and Computer Engineering, and IREAP, University of Maryland, College Park, MD 20742, USA}

\author{Mohammad Hafezi}
\affiliation{Joint Quantum Institute, NIST/University of Maryland, College Park, MD 20742, USA}
\affiliation{Department of Electrical and Computer Engineering, and IREAP, University of Maryland, College Park, MD 20742, USA}
\affiliation{Department of Physics, University of Maryland, College Park, MD 20742, USA}


\begin{abstract}
Recent advances in realizing optical frequency combs using nonlinear parametric processes in integrated photonic resonators have revolutionized on-chip optical clocks, spectroscopy, and multi-channel optical communications. At the same time, the introduction of topological physics in photonic systems has provided a new paradigm to engineer the flow of photons, and thereby, design photonic devices with novel functionalities and inherent robustness against fabrication disorders. Here, we use topological design principles to theoretically propose the generation of optical frequency combs and temporal Kerr solitons in a two-dimensional array of coupled ring resonators that creates a synthetic magnetic field for photons and exhibits topological edge states. We show that these topological edge states constitute a travelling-wave super-ring resonator that leads to the generation of coherent nested optical frequency combs, and self-formation of nested temporal solitons and Turing rolls that are remarkably phase-locked over $>$40 rings. In the nested soliton regime, our system operates as a pulsed optical frequency comb and achieves a mode efficiency of $>50\%$, an order of magnitude higher than single ring frequency combs that are theoretically limited to only $\sim 5\%$. Furthermore, we show that the topological nested solitons are robust against defects in the lattice. This topological frequency comb works in a parameter regime that can be readily accessed using existing low-loss integrated photonic platforms like silicon-nitride. Our results could pave the way for efficient on-chip optical frequency combs, and investigations of various other soliton solutions in conjunction with synthetic gauge fields and topological phenomena in large arrays of coupled resonators.
\end{abstract}

\maketitle

Optical frequency combs are a set of equidistant spectral lines that find a multitude of applications in metrology, spectroscopy, and precision clocks \cite{Udem2002, Cundiff2003, Kippenberg2011, Kippenberg2018, Pasquazi2018, Gaeta2019, Diddams2020, Suh2016}.  While optical frequency combs emerge naturally in mode-locked ultra-fast lasers \cite{Udem2002, Cundiff2003, Diddams2020}, using nonlinear parametric processes, in particular the Kerr effect, in integrated photonic resonators offers a much more convenient and compact route to generate optical frequency combs \cite{Kippenberg2011, Kippenberg2018, Pasquazi2018, Gaeta2019, DelHaye2007}. Integrated optical frequency combs have opened the route, for example, to on-chip transceivers for tera-bit scale wavelength-multiplexed optical communications \cite{Marin-Palomo2017}, chip-scale light detection and ranging (LiDAR) \cite{Riemensberger2020}, on-chip frequency synthesizers \cite{Spencer2018} and optical clocks \cite{Newman2019}. Of particular significance is the regime of coherent optical frequency combs where the intrinsic dispersion and dissipation of a photonic resonator are counterbalanced by the nonlinearity induced dispersion and parametric gain, respectively, and it leads to self-formation of stationary solutions called dissipative Kerr solitons (DKS) \cite{Kippenberg2018}. DKS have been demonstrated in a variety of single resonator geometries, and diverse material platforms such as silica glass, silicon nitride, etc. \cite{Kippenberg2018, Pasquazi2018, Herr2013}. More recently, DKS have been explored in photonic molecules, that is, a configuration of two coupled resonators, which allows exploration of collective coherence or self-organization, and also solitonic solutions that are inaccessible using single resonator \cite{Jang2018, Tikan2020, Helgason2020}.

In parallel, advances in the field of topological photonics have allowed access to new paradigms that can be used to design photonic devices with novel functionalities \cite{Lu2014, Khanikaev2017, Ozawa2019}. On one hand, topological photonic systems use complex arrays of hundreds of coupled waveguides or ring resonators \cite{Rechtsman2013, Hafezi2013}. On the other hand, such systems exhibit remarkably simple features, such as edge states, that are dictated only by the global topology and are, therefore, independent of local details of the system. This unique property of edge states protects them against local defects and disorders in the system, and has enabled the realization of robust photonic devices such as optical delay lines \cite{Hafezi2011, Hafezi2013, Mittal2014}, lasers \cite{StJean2017, Bahari2017, Bandres2018, Yang2020}, switches \cite{Cheng2016, Zhao2019}, photonic crystal waveguides and cavities \cite{Barik2018, Shalaev2019, Gao2019}, fibers \cite{Lu2018}, etc. Lately, topological edge states have also been used in conjunction with nonlinear parametric processes for efficient and tunable generation of quantum states of light via spontaneous four-wave mixing \cite{Mittal2018, Orre2020, Blanco-Redondo2018}, optical frequency conversion \cite{Kruk2019, Smirnova2020}, nonlinear optical switching \cite{Leykam2018}, and also to investigate spatial solitons in coupled waveguide arrays \cite{Lumer2013, Ablowitz2014, Leykam2016, Mukherjee2020, Marzuola2019}.

\begin{figure*}[!ht]
 \centering
 \includegraphics[width=0.98\textwidth]{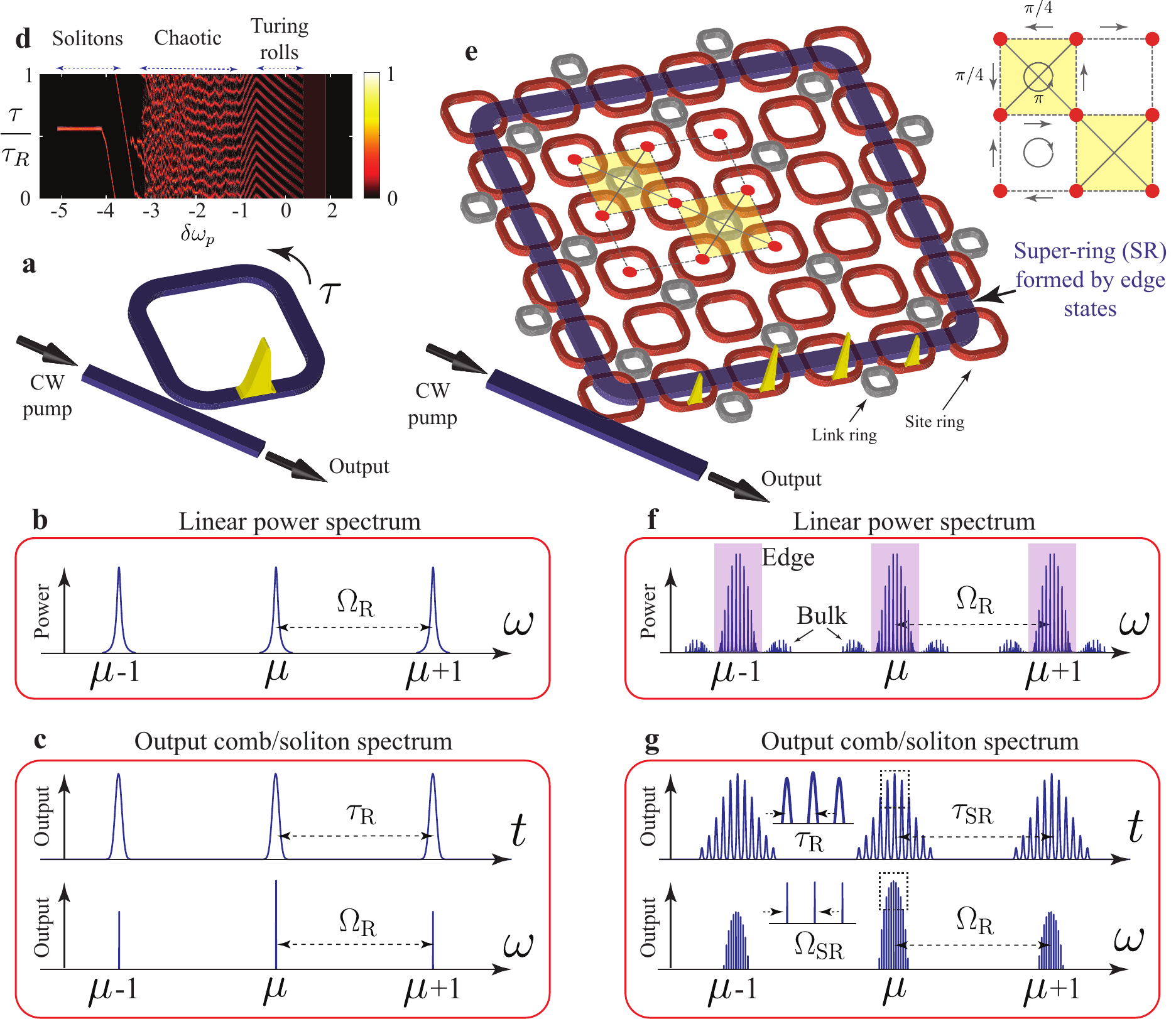}
 \caption {
 \textbf{a} Schematic of a single-ring resonator, and \textbf{b} power spectrum in the linear regime. The plot shows three longitudinal mode resonances, indexed by an integer $\mu$, and separated by a free spectral range (FSR) $\OR$. To generate optical frequency combs and temporal Kerr solitons, the ring resonator is pumped using a continuous-wave (CW) laser near one of the longitudinal mode resonances.
\textbf{c} Temporal and spectral response at the output of the ring resonator in the regime of Kerr solitons. The temporal output consists of a series of pulses separated by $\tau_{\text{R}}$, the round trip time of the ring resonator. The spectral output consists of a series of narrow lines separated by FSR $\OR$.
\textbf{d} An indicative spatio-temporal intensity distribution in the ring as a function of input pump frequency detuning $\delta\w_{p}$ from the cold-cavity resonance, normalized by the bandwidth of the resonator. The frequency comb operates in different regimes depending on the input pump frequency, as highlighted in the figure (also see Fig.S2 of the Supplementary Information). Notably, close to the cold-cavity resonance, the ring exhibits Turing rolls that are a series of equidistant pulses. Following this, there is a broad chaotic region where there are no well-defined features in the ring intensity. At higher pump frequency detunings, $\delta\w_{p} \sim -4$, the ring exhibits single or multiple Kerr solitons.
\textbf{e} Schematic of a 2D array of ring resonators that simulates the anomalous quantum-Hall effect for photons and exhibits topological edge modes at its boundary. The link rings are used to achieve a local synthetic magnetic field for photons, as shown in the inset.
\textbf{f} Power spectrum in the linear regime showing edge state resonances (shaded) and bulk bands. The edge states extend through out the boundary of the lattice and constitute a super-ring resonator, with longitudinal mode separation $\OSR$. The transmission spectrum repeats every FSR $\Omega_{\text{R}}$ of the ring resonators. To generate optical frequency comb, the topological super-ring is pumped by a CW laser near one of the edge mode resonances. The array can now host super solitons with spatial intensity profile shown by yellow colored pulses.
\textbf{g} Temporal and spectral response at the output of the topological frequency comb in the single soliton regime. The output temporal profile now consists of a series of soliton pulses separated by single ring round trip time $\TR$ (fast time), modulated by a series of super-soliton pulses separated by the round trip time $\TSR$ (slow time) of the edge mode super ring. The output spectral profile shows a series of edge mode resonances (longitudinal modes of the super ring, slow frequency) oscillating in each FSR (fast frequency) of the the single ring resonators.
}
 \label{fig:1}
\end{figure*}

Here, we theoretically propose the generation of optical frequency combs and temporal dissipative Kerr super-solitons in a topological photonic system consisting of a two-dimensional array of coupled (micro)ring resonators. This array creates a local synthetic magnetic field for photons, and thereby, exhibits photonic topological edge states that circulate around the boundary of the lattice \cite{Hafezi2011,Hafezi2013}. In our system, the edge states constitute a travelling-wave super-ring resonator formed of multiple single ring resonators (Fig.\ref{fig:1}\textbf{e}). We show that continuous-wave pumping of edge states leads to self-formation of temporal patterns, in particular, Turing rolls and dissipative Kerr super-solitons (DKSS) in the super-ring resonator formed by edge states. More importantly, we show that these temporal features are phase-locked across all the ring resonators on the edge of the lattice, indicating collective coherence or self-organization across $>$40 oscillators. Specifically, at low pump powers, we observe Turing rolls in each resonator that forms the edge state super-ring resonator, and the phase of the Turing rolls in each resonator on the edge is locked. At higher pump powers, we demonstrate the generation of temporal nested-solitons - a configuration where we observe super-soliton formation in the super-ring edge resonator, and solitons within each ring resonator that is a part of this super-soliton. Remarkably, we observe that the phases of the solitons in individual ring resonators are locked as this super-soliton structure circulates around the edge of the lattice. This collective-coherence of $>$40 rings in a 2D array of $>$100 rings is enabled by the presence of topological edge states in our device that constrain multiple ring resonators to oscillate in the form of a super-ring resonator. Therefore, from a fundamental perspective, our results open the route to engineer nonlinear parametric processes, self-organization, and temporal soliton formation using synthetic magnetic fields and topological principles.

From an application perspective, in the super-soliton regime, the frequency and the temporal spectra of our topological system correspond to that of a pulsed optical frequency comb \cite{Bao2019,Xue2019}. This allows our topological frequency comb to achieve a mode efficiency of $>50\%$, an order of magnitude higher than that observed in single ring frequency combs \cite{Xue2017,Bao2019,Xue2019}. Furthermore, we show that the super-solitons formed by edge states inherit the topological protection of a linear system and are robust against any defects in the lattice. Our design can be implemented readily with the existing nanofabrication technology, and similar topological ring resonator systems with nonlinear parametric processes have already been realized to enhance and engineer the generation of quantum states of light \cite{Mittal2018, Orre2020}. Therefore, our results could pave the way for the on-chip realization of efficient, and topologically robust pulsed optical frequency combs and dissipative Kerr super-solitons.

In the following, we will first describe our topological system and the model that we use to simulate the linear and nonlinear processes in the system. We will then show the emergence of self-organized Turing rolls at low pump powers. Following this, at higher pump powers, we will demonstrate the generation of super-solitons in the lattice that achieve higher mode efficiency compared to single-ring frequency combs and are topologically robust. We will conclude with a discussion of the phase diagram and the experimental implementation of the topological frequency comb, and our outlook for extensions of frequency combs to other coupled-resonator topological systems and lattice models. \\


\noindent
\textbf{The topological system and simulation framework}

Our topological system consists of an array of ring resonators located at the sites of a square lattice (Fig.\ref{fig:1}\textbf{e}). These site ring resonators are coupled to their nearest and next-nearest neighbors using another set of rings, which we call the link rings \cite{Leykam2018,Mittal2019}. The longitudinal mode resonance frequencies of the link rings are detuned from those of the site rings by introducing a path length difference such that the detuning is one half free-spectral range (FSR, $\OR$), which is the frequency separation between consecutive longitudinal modes of the ring resonators \cite{Hafezi2011, Hafezi2013, Mittal2019}. Therefore, near site-ring resonance frequencies, light intensity in the link rings is negligible. As such, the link rings act as waveguides and introduce a non-zero local synthetic magnetic field such that the flux through a half-unit cell is $\pi$ but the total flux through a unit cell is zero (see inset of Fig.\ref{fig:1}\textbf{e}). This lattice simulates the anomalous quantum Hall model for photons, and in the linear regime, the dynamics close to a longitudinal mode resonance of the site rings is described by a tight-binding Hamiltonian which is similar to that of the Haldane model \cite{Haldane1988}
\begin{eqnarray} \label{Eq_HL}
H_{L} &=& \sum_{m,\mu} \w_{0,\mu} \ad_{m,\mu} \a_{m,\mu} - J \sum_{\left<m,n\right>} \ad_{m,\mu} ~\a_{n,\mu} e^{-i \phi_{m,n}}  \nonumber \\
      &-&  J \sum_{\left<\left<m,n\right>\right>} \ad_{m,\mu} ~\a_{m,\mu} + \mathrm{h.c.}
\end{eqnarray}
Here $a_{m,\mu}$ is the photon annihilation operator at a (site-ring) spatial position $m = \left(x,y\right)$, and $\omega_{0,\mu}$ is the resonance frequency of a site ring resonator longitudinal mode with the integer index $\mu$. $J$ is the coupling strength between ring resonators, and is the same for both nearest (indicated by $\left<m,n \right>$) and next-nearest neighbor (indicated by $\left<\left<m,n\right>\right>$) couplings. The hopping phase $\phi_{m,n} = \pm \frac{\pi}{4}$ for nearest-neighbor couplings and is zero for the next-nearest neighbor couplings. Note that this tight-binding description of our system is valid in the regime $J \ll \OR$, and in this linear Hamiltonian the photon operators in different longitudinal modes (indexed by $\mu$) are uncoupled. We also emphasize that our system is time-reversal symmetric and supports two pseudo-spins that correspond to the circulation direction of photons within each site-ring resonators \cite{Hafezi2011, Hafezi2013, Leykam2018, Mittal2019}, and is therefore, analogous to the quantum spin-Hall system. Nevertheless, it has been experimentally demonstrated that we can excite a single pseudo-spin in the system, and the parasitic coupling between the two pseudo-spins is negligible such that it effectively breaks time-reversal symmetry for a given pseudo-spin \cite{Hafezi2013, Mittal2014, Mittal2019}.

To probe this lattice, we couple one ring at the edge of the lattice to input-output waveguide as shown in Fig.\ref{fig:1}\textbf{e}. The transmission spectrum of the lattice exhibits two bulk bands separated by a band $\left( \left[-1, 1\right] J \right)$ that hosts topological edge states (Fig.\ref{fig:1}\textbf{f}, Supplementary Fig.S1). The edge states propagate along the boundary of the lattice and therefore, constitute a travelling-wave super-ring resonator (see Fig.\ref{fig:1}\textbf{e} and Supplementary Fig.S1). The multiple edge state resonances in the edge band are then the longitudinal modes of the super-ring, separated by a free spectral range $\OSR$. The bulk state wavefunctions, in contrast, occupy the bulk of the lattice and do not have a well-defined propagation path in the lattice. Moreover, the edge states are characterized by topologically invariant integers, and are therefore, remarkably robust against defects and disorders in the lattice \cite{Ozawa2019, Hafezi2011, Hafezi2013, Mittal2014, Mittal2019}. Furthermore, this structure of an edge band sandwiched between two bulk bands repeats in each FSR ($\OR$) of the individual ring resonators. The super-ring resonator formed by the edge states can be compared to that of a single ring resonator all-pass filter (APF, Fig.\ref{fig:1}\textbf{a}) where each longitudinal mode resonance of the APF is now split into a number of super-modes (Fig.\ref{fig:1}\textbf{f}).

To generate optical frequency comb in this lattice, we inject a continuous-wave (CW) pump laser with frequency close to one of the longitudinal mode resonances $\left(\mu = 0 \right)$ using the input-output waveguide (Fig.\ref{fig:1}\textbf{e}). The intrinsic Kerr nonlinearity of the ring resonators leads to spontaneous four-wave mixing, and subsequently, stimulated generation of photons in other longitudinal modes $\left(\mu \neq 0 \right)$. This nonlinear Kerr interaction is described by the Hamiltonian
\begin{equation} \label{Eq_HNL}
H_{NL} = - \gamma ~\sum_{m,\mu}   \ad_{m,\mu_{1}} ~ \ad_{m,\mu_{2}} ~ \a_{m,\mu_{3}} ~\a_{m,\mu_{4}} \delta_{\mu_{1}+\mu_{2},\mu_{3}+\mu_{4}},
\end{equation}
where two photons at longitudinal modes $\mu_{1}$ and $\mu_{2}$ annihilate and generate two photons at longitudinal modes $\mu_{3}$ and $\mu_{4}$. $\delta_{\mu_{1}+\mu_{2},\mu_{3}+\mu_{4}}$ indicates conservation of energy and momentum in the four-wave mixing interaction, and $\gamma$ is the strength of nonlinear interaction. Note that the nonlinear Kerr interaction is local to a ring resonator, and couples multiple super-ring (edge state) resonances. Therefore, as shown in Fig.\ref{fig:1}\textbf{g}, in our topological frequency comb we intuitively expect oscillation of multiple edge modes within each longitudinal mode of the individual ring resonators. The expected temporal profile at the output then corresponds to a pulsed optical frequency comb, with bursts of pulses separated by a time interval $\TSR$, the round-trip time of the super-ring resonator (Fig.\ref{fig:1}\textbf{g}). The time interval between each pulse is $\TR$, the round trip time of a single ring resonator.

\begin{figure*}
 \centering
 \includegraphics[width=0.98\textwidth]{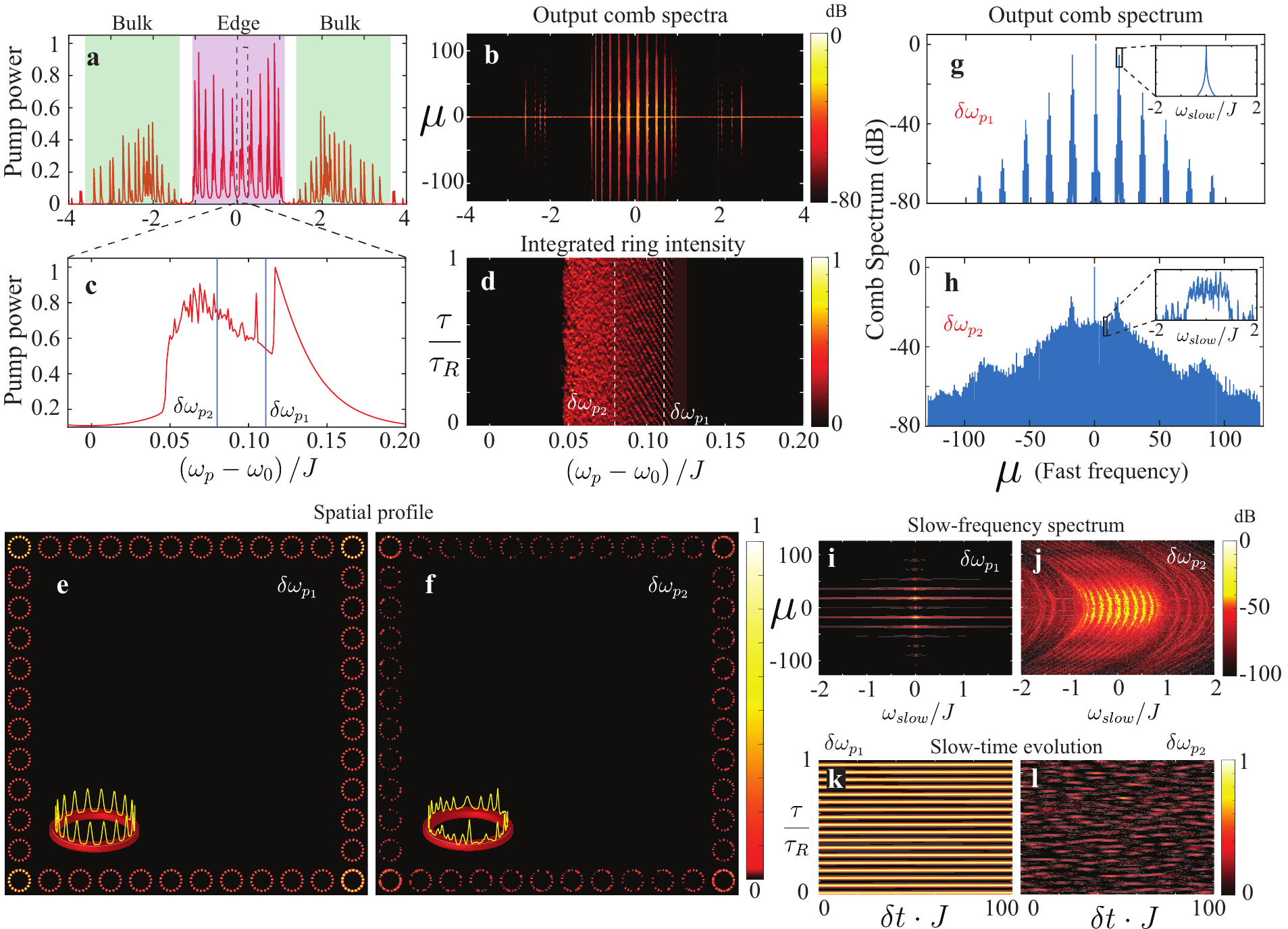}
 \caption{
 \textbf{a} Total pump power in the super-ring resonator as a function of the input pump frequency detuning $\left(\omega_{p}-\omega_{0}\right)/J$. Here $\omega_{p}$ in the input pump frequency, $\omega_{0}$ is the cold-cavity resonance of the pumped FSR (the longitudinal modes of single ring resonators), and $J$ is the coupling strength between resonators. The normalized input pump field $\E = 1.1$.
 \textbf{b} Spectral power of the generated frequency comb in different FSRs (indexed by $\mu$, the fast frequency) as a function of the input pump frequency. We observe that only the edge mode resonances lead to a significant generation of power in the optical frequency comb.
 \textbf{c} Total pump power in the super-ring resonator, for pump frequencies in one of the edge state resonances.
 \textbf{d} Spatio-temporal (or fast-time) intensity distribution in the ring resonators, integrated over the rings on the edge of the lattice, as a function of the pump frequency detuning. This plot can be compared with that of a single ring resonator, shown in the inset of Fig.1\textbf{a}. We analyze two different pump frequencies, $\delta\omega_{p_1} = 0.111 J$ and $\delta \omega_{p_2} = 0.08 J$ as indicated in \textbf{c} and \textbf{d}.
 \textbf{e,f} Spatial intensity distribution in the lattice,
 \textbf{g,h} output comb spectrum at $\delta\omega_{p_1}$ and $\delta\omega_{p_2}$, respectively. At $\delta\omega_{p_1}$ we observe phase-locked Turing rolls in each ring throughout the edge. The number of Turing rolls (18) in each ring corresponds to the difference in the oscillating FSRs (the fast-frequency, see \textbf{g}, \textbf{i}). At $\delta\omega_{p_2}$ the spatial intensity distribution in the rings is no longer locked and the comb spectrum  suggests a chaotic regime. Note that the comb intensity in the bulk of the lattice is negligible even in the chaotic regime.
\textbf{i,j} Slow frequency spectrum of the output comb at $\delta\omega_{p_1}$ and $\delta\omega_{p_2}$, respectively. At $\delta\omega_{p_1}$, that is, in the region of Turing rolls, a single edge mode is oscillating (also see inset of \textbf{g}) in the bright FSRs. In chaotic regime, at $\delta\omega_{p_2}$, many edge modes are oscillating (inset of \textbf{h}). The underlying quadratic dispersion of the ring resonators is also visible.
\textbf{k,l} Corresponding slow-time temporal profile at the output. At $\delta\omega_{p_1}$ there is no long-term dynamics of the intensity distribution, consistent with the observation of a single edge mode oscillating in \textbf{i}. At $\delta\omega_{p_2}$, the temporal evolution is random which indicates oscillation of multiple edge mode resonances without any phase coherence.
 }
 \label{fig:2}
\end{figure*}

To simulate our topological frequency comb we use the linear and nonlinear tight-binding Hamiltonians of Eqs.(\ref{Eq_HL}), (\ref{Eq_HNL}) and derive the coupled driven-dissipative nonlinear Schr\"{o}dinger equations, also called Lugiato-Lefever equations (LLEs) \cite{Chembo2010, Chembo2013, Pasquazi2018}, that dictate the spatial, spectral and temporal evolution of site ring amplitudes as
\begin{eqnarray} \label{Eq_Diff}
\frac{da_{m,\mu}}{dt} &=& -i \left(\w_{0,\mu} - \wp - \OR ~\mu \right) \a_{m,\mu} - J \sum_{\left<n\right>} \a_{n,\mu} e^{-i \phi_{m,n}} \nonumber \\
                      &-&  J \sum_{\left<\left<n\right>\right>} \a_{n,\mu} + i \gamma ~\frac{1}{\TR} \int_{0}^{\TR} d\tau \left( \left|a_{m,\tau} \right|^{2} a_{m,\tau} \right) e^{-i \w_{\mu} \tau} \nonumber \\
                      &-& \left(\kex ~\delta_{m,\text{IO}} + \kin \right) a_{m,\mu} + \delta_{m,\text{IO}} ~\delta_{\mu,0} ~\E.
\end{eqnarray}
Here, the coupling between nearest and next-nearest neighbors is expressed in longitudinal mode resonances (indexed by integer $\mu$) of the ring resonators. We include second-order anomalous dispersion $D_{2}$ in the longitudinal mode resonances such that $\omega_{0,\mu} = \wo + \OR ~\mu + \frac{D_{2}}{2} ~\mu^2$. The input pump frequency is denoted by $\wp$ and the pumped longitudinal mode corresponds to $\mu = 0$ with resonance frequency $\wo$. Therefore, in a reference frame that is rotating at frequency $\OR$, $\left(\w_{0,\mu} - \wp - \OR ~\mu \right)$ is the effective detuning of the pump frequency from respective longitudinal mode resonances (indexed by the integer $\mu$). $\kex$ is the coupling rate of the input-output ring (at the corner) to the external waveguide, and $\kin$ is the loss rate of the ring resonators. $\E$ is the normalized input pump field which is coupled only to the input-output ($\text{IO}$) ring, and in the longitudinal mode $\mu = 0$.

The nonlinear four-wave mixing interaction between different longitudinal mode resonances (\ref{Eq_HNL}) is represented in the time domain $\tau$ in $\left[0, \TR \right]$ which corresponds to the round-trip time within a single ring resonator \cite{Hansson2014, Chembo2014}. In a reference frame rotating at frequency $\OR = \frac{2\pi}{\TR}$, $a_{m,\tau}$ represents the spatio-temporal field within a ring, at a lattice location $m = \left(x,y\right)$, and is related to the spectral field within a ring as
\begin{equation}\label{Eq_FT}
a_{m,\mu} = \frac{1}{\TR} \int_{0}^{\TR} d\tau ~a_{m,\tau} e^{-i \w_{\mu} \tau}.
\end{equation}

The spectral dynamics of our system involves two frequency scales, the longitudinal mode resonances $\sim \mu \OR$ of the individual ring resonators and the coupling strength $J$ between resonators. Because $J \ll \OR$, we refer to the longitudinal mode resonances (indexed by integer $\mu$) of the individual ring resonators as the fast frequency. The frequency response of our system near each longitudinal mode resonance $\mu$ is dictated by the coupling strength $J$ (see Eq.\ref{Eq_HL} and Supplementary Information Fig.S1), and we refer to it as the slow frequency $\left(\omega_{slow} \right)$. Correspondingly, the temporal dynamics of our system also involves two time scales, the fast-time $\tau$ and the slow-time $t$. As we mentioned earlier, the fast-time $\tau$ represents the spatio-temporal field distribution within a ring resonator and is related to the fast-frequency $\mu$ via Eq.\ref{Eq_FT}. The slow-time $t$ depicts the evolution of spatio-temporal field distributions $\left(a_{m,\tau}\right)$, or equivalently the fast-frequency components $\left(a_{m,\mu}\right)$, at a time scale $\sim 1/J$. Therefore, the Fourier-transform of the slow-time evolution of $a_{m,\mu}$ yields the slow-frequency $\left(\omega_{slow} \right)$ spectrum, that is, the spectral information within a given longitudinal mode $\mu$. Thus coupled Eqs.(\ref{Eq_Diff}) reveal the complete spatial, spectral and temporal dynamics of our system. Note that we have not made any assumptions regarding the spectral position, bandwidth, or dispersion of the edge state resonances with in a longitudinal mode $\mu$ (see Supplementary Information, Section VI). Nevertheless, for simplicity, we have assumed that the coupling rate, the loss rate, and the hopping phase are same for all longitudinal modes.

For our numerical simulations, we consider a $12 \times 12$ lattice of site rings. We use dimensionless units \cite{Chembo2010, Chembo2013}, and normalize parameters such that $J = 1$, $\gamma = 1$, $\kex = 0.05$, $\kin = 0.005$, and $D_{2} = 0.0025$. The coupling rate $\kex$ and the loss rate $\kin$ are chosen such that the individual edge state resonances in the edge band are resolved. We consider 256 FSRs of the individual ring resonators. Because the coupled Eqs.(\ref{Eq_Diff}) are written in a reference frame rotating at frequency $\OR$, the FSR of the individual ring resonators is an independent parameter, as is chosen to be $\OR = 20$.\\

\noindent
\textbf{Turing Rolls and Collective-Coherence}

To understand the generation of optical frequency combs in our topological device, we first fix the (normalized) input pump field at $\E = 1.1$, and observe the output spectra of generated photons across multiple FSRs (fast-frequency, indexed by $\mu$) as we tune the input pump frequency in one of the FSRs (Fig.\ref{fig:2}\textbf{a,b}). The pump frequency is denoted by the detuning $\delta\omega_{p} = \frac{\left(\omega_{p} - \omega_{0}\right)}{J}$, where $\omega_{0}$ is the cold cavity resonance of the pumped FSR $\left( \mu = 0 \right)$. We find that the generation of frequency comb (bright light intensity across multiple FSRs) is efficient only when the input pump frequency is close to one of the edge mode resonances. Furthermore, as we will show in the following, on pumping near edge mode resonances, the bright frequency comb is generated only in the ring resonators that lie on the edge of the lattice. By contrast, when the input pump frequency is in the bulk bands, the generation of light in FSRs other than the pumped FSR is very weak and the frequency comb is inefficient. We recall that, in the linear regime, the edge state resonances constitute a travelling-wave super-ring resonator, whereas, the bulk states do not have a well-defined momentum in the lattice (see Supplementary Information, Fig.S1). We find that the edge states preserve their characteristics even in the presence of nonlinearities, and therefore, the super-ring resonator formed by the edge states efficiently reinforces the optical frequency comb. This observation is also consistent with experimental observations on a similar nonlinear topological ring resonator system that show edge states lead to enhancement of spontaneous-four wave mixing and generation of energy-time entangled photon pairs \cite{Mittal2018, Orre2020}.

Having established that the topological optical frequency comb is efficient only when the input pump frequencies excite the edge modes, we now focus on pump frequencies near a single edge mode of the lattice. The pump power in the input-output ring resonator is shown in Fig.\ref{fig:2}\textbf{c} as a function of the pump frequency detuning from $\w_{0}$, the cold cavity resonance of the pumped FSR (of single ring resonators). In Fig.\ref{fig:2}\textbf{d}, we plot $\sum_{m \in \text{edge}} abs\left|a_{m,\tau} \right|^2$, that is, the spatio-temporal intensity distribution (or equivalently, the fast-time $\tau$ distribution) integrated over the rings on the edge of the lattice as a function of the input pump frequency. The presence of sharp features in this plot indicates both, the self-formation of temporal (or equivalently, spatial) features within individual ring resonators and the self-organization of these features across rings. Randomly varying features in this plot indicate randomness in the spatial intensity distribution within rings or a lack of coherence between rings. This plot can be compared with the corresponding plot of a single ring resonator frequency comb shown in Fig.\ref{fig:1}\textbf{d} (also see Supplementary Information Fig.S2).

When the input pump frequency is at $\delta\omega_{p_{1}} = 0.111 J$, in Fig.\ref{fig:2}\textbf{d} we observe a regularly oscillating pattern along the fast-time $\tau$ axis that indicates formation of Turing rolls \cite{Pasquazi2018, Kippenberg2018} (also see Supplementary Fig.S2). To confirm, in Fig.\ref{fig:2}\textbf{e} we plot the spatio-temporal intensity distribution within each ring of the lattice, that is, $\left|a_{m,\tau} \right|^2$ where $m = \left(x,y\right)$ indicates location of a ring in the lattice. Indeed, we find that all the intensity distribution in all the rings on the edge of the lattice exhibit equally spaced pulses, called Turing rolls \cite{Pasquazi2018, Kippenberg2018}, and the intensity in the bulk of the lattice is negligible. More remarkably, we find that the phase of the Turing rolls is locked throughout the edge of the lattice. This shows self-organization or collective spectro-temporal coherence between all the 44 rings on the edge. In fact, we find a broad region, of bandwidth $\approx 0.01 J$, where we observe coherent Turing rolls. From Fig.\ref{fig:2}\textbf{c}, we also see that the pump power in the input-output ring varies smoothly in this region of coherent Turing rolls (also see Supplementary Information Fig.S2). Note that the light intensity on the corner rings of the lattice is higher than that on the edge of the lattice because the corners act as defects in the lattice (also see Supplementary Information Fig.S1 \textbf{e}). Nevertheless, because the edge states are topologically protected, the corners can not scatter light backward or into the bulk of the lattice.

For lower pump frequencies (near $\delta\omega_{p_{2}}$ in Fig.\ref{fig:2}\textbf{c,d}), we find a chaotic region  where the pump power varies rapidly as we tune the input pump frequency. More importantly, we see that spatial intensity distribution within each ring, and also the distribution across rings, is now random without any coherence. Nevertheless, we observe that the comb intensity is still confined to the edge of the lattice and the intensity in the bulk is negligible.

We now discuss the spectral and temporal distributions at the output of our topological frequency comb. The output comb spectra for the two pump frequencies $\delta\omega_{p_{1}}$ and $\delta\omega_{p_{2}}$ are shown in Figs.\ref{fig:2}\textbf{g,h}. For $\delta\omega_{p_{1}}$, the frequency comb spectrum predominantly consists of discrete spectral lines, separated by 18 FSRs. As in the case of a single-ring frequency comb, this number corresponds to exactly the number of Turing rolls in each ring (Fig.\ref{fig:2}\textbf{e}). Furthermore, we have confirmed that the number of Turing rolls decreases as $\frac{1}{\sqrt{D_{2}}}$, as is the case for a single-ring optical frequency comb \cite{Pasquazi2018, Chembo2010, Chembo2014} (see Supplementary Information Fig.S3). At $\delta\omega_{p_{2}}$, that is, in the chaotic region, the discrete lines in the primary comb have merged together, and there are no distinct features in the frequency spectrum.

As we mentioned earlier, the slow-time $\left(t \right)$ dynamics of coupled Eqs.(\ref{Eq_Diff}) allows us to reconstruct the slow-frequency $\omega_{slow}$ spectrum, that is, frequencies of the order of $J$ within each longitudinal mode (indexed by integer $\mu$, the fast frequency) of the individual ring resonators. However, because $\omega_{slow} \approx J \ll \OR$, the slow-frequency spectrum is not resolved in Figs.\ref{fig:2}\textbf{g},\textbf{h} that show spectra along the fast-frequency axis (FSRs, $\mu$). Therefore, to better visualize the slow-frequency response of our topological comb, in Figs.\ref{fig:2}\textbf{i},\textbf{j} we show a 2D plot of the slow-frequency spectrum with in each FSR. Here, the slow-frequency $\omega_{slow}$ ($x-$axis) is calculated as the detuning from the corresponding longitudinal mode resonance frequency $\omega_{0,\mu}$ and the input pump frequency, such that  $\omega_{slow} = \left\{ \left(\omega_{\mu} - \omega_{0,\mu}\right) - \left(\omega_{p} - \omega_{0}\right) \right\} / J$. $\omega_{\mu}$ is the frequency of generated light in a longitudinal mode $\mu$.

At $\delta\omega_{p_{1}}$, that is, in the region of coherent Turing rolls, the slow-frequency spectrum within each bright FSR (the fast frequency, $\mu$) exhibits a single mode centered around the comb line (see inset of Fig.\ref{fig:2} \textbf{g} for a cross-section of this plot). The oscillation of a single edge mode within each oscillating FSR is consistent with the uniform spatial intensity distribution that we observed in Fig.\ref{fig:2}\textbf{e}. This can also be inferred from the slow-time ($t$) evolution of the fast-time $\tau$ intensity distribution in the input-output ring which is a constant.

At $\delta\omega_{p_{2}}$, that is, in the chaotic region, we observe oscillation of multiple modes in the edge band ($\omega_{slow} = \left(-1,1\right) J$) of each FSR. This oscillation of multiple edge modes in the chaotic regime is also evident from the non-uniformity of the spatial intensity distribution in the lattice (Fig.\ref{fig:2}\textbf{f}), and the dynamics of the output temporal profile (Fig.\ref{fig:2}\textbf{l}) which varies with slow-time $t$. Furthermore, spectral power in bulk modes ($\omega_{slow} < -1J, ~\text{and} ~\omega_{slow} > 1J$) is two orders of magnitude smaller compared to those of the edge modes. This validates our observation of negligible light intensity in the bulk of the lattice (Fig.\ref{fig:2} \textbf{f}). Note that the oscillating edge modes in Fig.\ref{fig:2}\textbf{j}) also show the underlying quadratic dispersion of the ring resonators in different FSRs. \\

\noindent
\textbf{Temporal Kerr Super-Solitons}

\begin{figure*}
 \centering
 \includegraphics[width=0.98\textwidth]{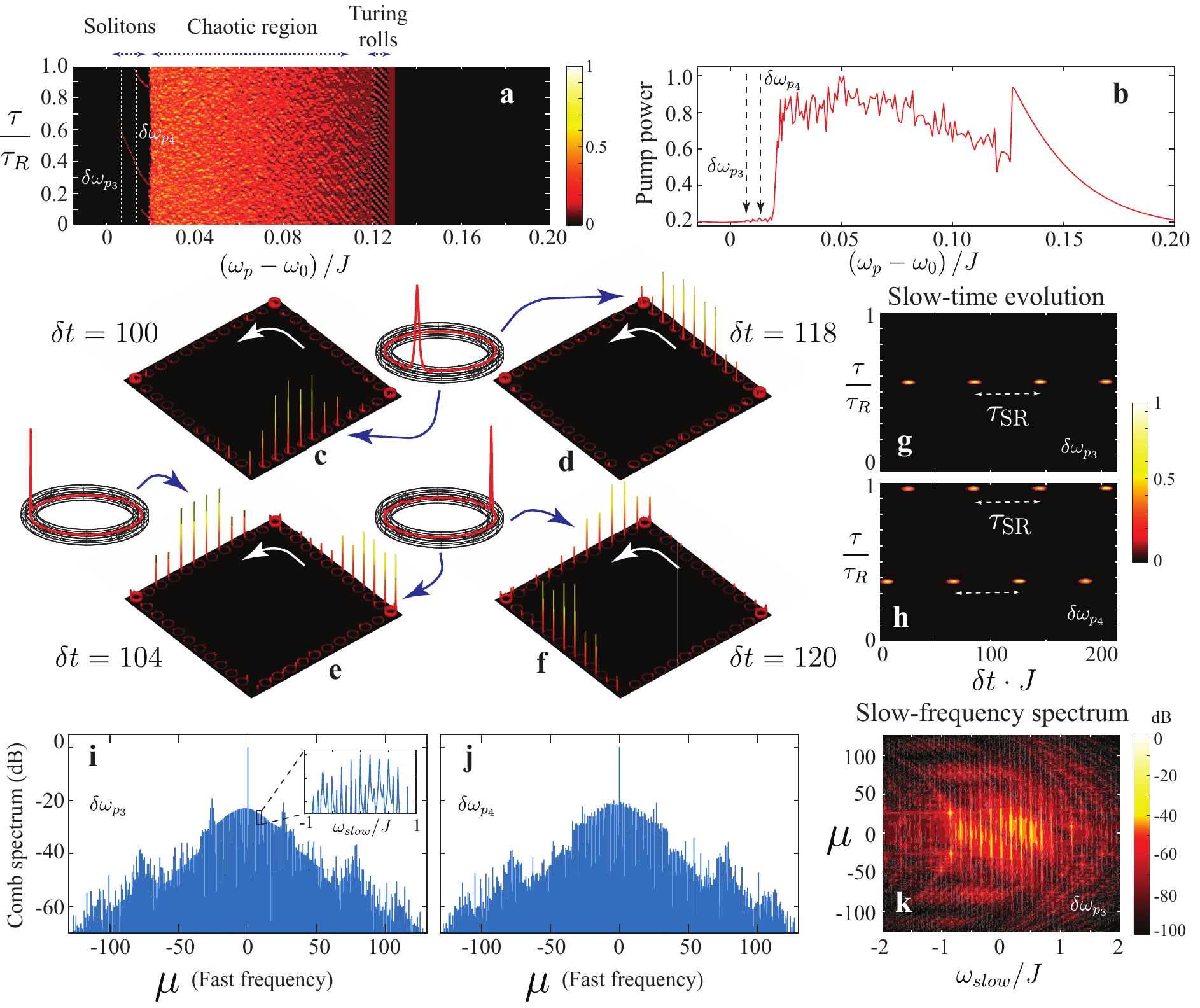}
 \caption {
 \textbf{a} Spatio-temporal (or fast-time) intensity distribution as a function of input pump frequency, integrated over edge rings. We analyze two different input pump frequencies, $\delta\omega_{p_3}$ and $\delta\omega_{p_4}$.
 \textbf{b} Total pump power in the super-ring resonator, with input pump field $\E = 1.56$.
 \textbf{c,d} Spatial intensity profiles in the lattice at $\delta\omega_{p_3}$, for two different slow times. We observe super-solitons, that is, a soliton in the edge mode super-ring resonator which is formed of solitons in the individual ring resonators. The super-ring soliton circulates around the lattice in the counter-clockwise direction (see Movie M1). Moreover, the phase (or position) of the soliton in each ring resonator on the edge remains same as the super-soliton propagates around the edge.
\textbf{e,f} Spatial intensity profiles in the lattice at $\delta\omega_{p_4}$, for two different slow times. We now observe two solitons that also go around the edge in counter-clockwise direction (see Movie M2). However, the phase of the two solitons (see insets) is now different in the two solitons.
\textbf{g,h} Corresponding output temporal profiles. At $\delta\omega_{p_3}$ we observe super-soliton pulses with a repetition time of $\tau_{\text{SR}}$ - the round trip time of the edge mode super-ring. Note that each of these super-soliton pulses is composed on soliton pulses from individual ring resonators, separated by a much smaller delay $\tau_{\text{R}}$ (see Fig.\ref{fig:1}\textbf{g}). At $\delta\omega_{p_4}$, there are two super-soliton pulses in each round trip time $\tau_{\text{SR}}$.
\textbf{i,j} Output comb spectra for the two pump frequencies. At $\delta\omega_{p_3}$ the frequency spectrum is phase-locked. The inset shows slow-frequency spectrum where multiple equidistant edge modes are oscillating.
\textbf{k} Slow frequency spectrum at $\delta\omega_{p_3}$, showing oscillation of individual edge mode resonances with an effectively linear dispersion.
 }
 \label{fig:3}
\end{figure*}

We now proceed to show the presence of super-solitons in our topological frequency comb. We increase the normalized input pump field to $\E = 1.56$. As before, in Fig.\ref{fig:3}\textbf{a} we plot the spatio-temporal intensity distribution in ring resonators, integrated over the rings on the edge of the lattice, that is, $\sum_{m \in \text{edge}} abs\left|a_{m,\tau} \right|^2$. In addition to the coherent Turing rolls and chaotic regions, we observe a new region $\left(0 - 0.02 J\right)$ where light intensity is confined to very narrow regions (thin strands) in the ring resonators. A quick comparison with the analogous spatio-temporal intensity distribution of a single-ring optical frequency comb (see Fig.\ref{fig:1}\textbf{d}, and refs.\cite{Kippenberg2018, Pasquazi2018}) reveals that this region hosts solitons in our topological frequency comb. Figure \ref{fig:3}\textbf{b} shows the total pump power in the super-ring resonator. We see the emergence of kinks in the region where we expect solitons.

We analyze two different pump frequencies $\delta\omega_{p_{3}} = 0.0007 J$ and $\delta\omega_{p_{4}} = 0.0135 J$ in this region, as indicated in the figure. From the spatial intensity distribution (Fig.\ref{fig:3}\textbf{c}), we see that at $\delta\omega_{p_{3}}$, our topological frequency comb exhibits nested-solitons: the intensity distribution along the super-resonator (edge) of the lattice is confined to a small region of the edge in the form of a super-soliton, and the intensity distribution within each ring is also confined to a narrow region in the form of a soliton. This super-soliton then circulates along the edge of the lattice in a counter-clockwise direction as the slow-time $t$ evolves (see Movie M1). Remarkably, the spatio-temporal phase of the solitons in individual rings of the super-soliton is locked (see inset of Fig.\ref{fig:3}\textbf{c}) as this super-soliton structure circulates around the edge of the lattice. This observation once again highlights the collective coherence or self-organization of multiple nonlinear ring resonators on the edge of the lattice. The corresponding temporal spectrum (Fig.\ref{fig:3}\textbf{g}) at the output of our topological frequency comb then shows pulses of light that are separated by $\TSR$, the round-trip time of the super-resonator. Note that our simulations are carried out in a reference frame that is rotating at frequency $\Omega$, that is, co-propagating with the solitons in individual rings. Therefore, the solitons in each ring are actually circulating with time period $\TR$, and the super-soliton pulses in Fig.\ref{fig:3}\textbf{g} are actually bursts of pulses separated by time $\TR$ (see Fig.\ref{fig:1}\textbf{g}).

At input pump frequency $\delta\omega_{p_{4}}$, we observe two sets of super-solitons that are simultaneously circulating along the edge of the lattice (Fig.\ref{fig:3}\textbf{e,f}, Movie M2). Furthermore, while the phases of the individual ring solitons within each super-soliton are locked, the corresponding phases in the two super-solitons are not the same (insets of Fig.\ref{fig:3}\textbf{e,f}). In contrast to a single super-soliton, the temporal spectrum at the output of our topological frequency comb now consists of two bursts of pulses in each round trip time $\TSR$ of the super-ring resonator. At other pump frequencies in the super-soliton region, we can also observe three super-solitons.

Next, we discuss the frequency spectrum at the output of our topological frequency comb (Fig.\ref{fig:3}\textbf{i,j}). In the case of a single super-soliton, that is, at $\delta\omega_{p_{3}}$ the output frequency spectrum is, in general, smooth which indicates it is phase locked (except for few phase jumps, see Supplementary Information Fig.S4). By contrast, in the case of two super-solitons, that is, at $\delta\omega_{p_{4}}$ the frequency spectrum shows small variations (Fig.\ref{fig:3}\textbf{j}). This behaviour of the frequency spectra is similar to that observed in single ring resonator frequency combs where the spectrum is phase locked only when there exists a single soliton in the ring \cite{Chembo2010, Chembo2013, Pasquazi2018, Kippenberg2018}.

Furthermore, by resolving the slow-frequency ($\omega_{slow}$) response of our topological comb (Fig.\ref{fig:3}\textbf{k}), we find multiple edge modes that are oscillating within each FSR (also see inset of Fig.\ref{fig:3}\textbf{i}). More importantly, we see that the oscillating edge modes are equally-spaced within a given FSR and across FSRs, that is, the intrinsic (linear) dispersion of the longitudinal modes, of both the individual ring resonators and the super ring resonator, has now been exactly cancelled by the dispersion induced by Kerr nonlinearity. The slow-frequency response also explains the emergence of kinks in the Fig.\ref{fig:3}\textbf{i} - these are the regions where the dispersion curves from two different edge modes interfere and lead to phase jumps in the otherwise coherent frequency comb. This slow-frequency spectrum in the soliton regime can be compared to that of the chaotic regime (Fig.\ref{fig:2}\textbf{j}, Supplementary Information Fig. S4) where multiple modes in the edge band are oscillating but there is no cancellation of dispersion and no phase coherence.

A figure of merit for optical frequency combs is the mode efficiency at the output of the device, that is, the fraction of power that resides in the comb lines other than the pumped mode. For micro-ring resonator based optical frequency combs, the mode efficiency is limited to $\approx 5\%$ \cite{Xue2017,Bao2019,Xue2019}. This is mainly because the bright soliton coexists with a continuous-wave pump background. Pulsed optical frequency combs, implemented using two coupled cavities, one for the laser pump and the other for solitons, have been shown to alleviate this problem and achieve an efficiency of $75\%$ \cite{Bao2019,Xue2019}. However, these configurations require two cavities with disparate lengths, for example, a ring resonator and a fiber loop. Our topological device also implements a pulsed optical frequency comb as can be from its temporal and spectral response at the output. Nevertheless, in our device super-solitons in the super-ring edge resonator emerge naturally in a 2D array of identical coupled ring resonators because of its topological properties. More importantly, in the DKSS regime, we observe that $53\%$ of the total output power (in the waveguide) is contained in comb lines other than the pumped edge mode. This mode efficiency is an order of magnitude higher than that of single ring resonators, and can be achieved in a fully integrated photonics platform. We emphasize that the theoretical limit on the conversion efficiency of our topological frequency comb could be higher for other parameter regimes.

\begin{figure*}
 \centering
 \includegraphics[width=0.98\textwidth]{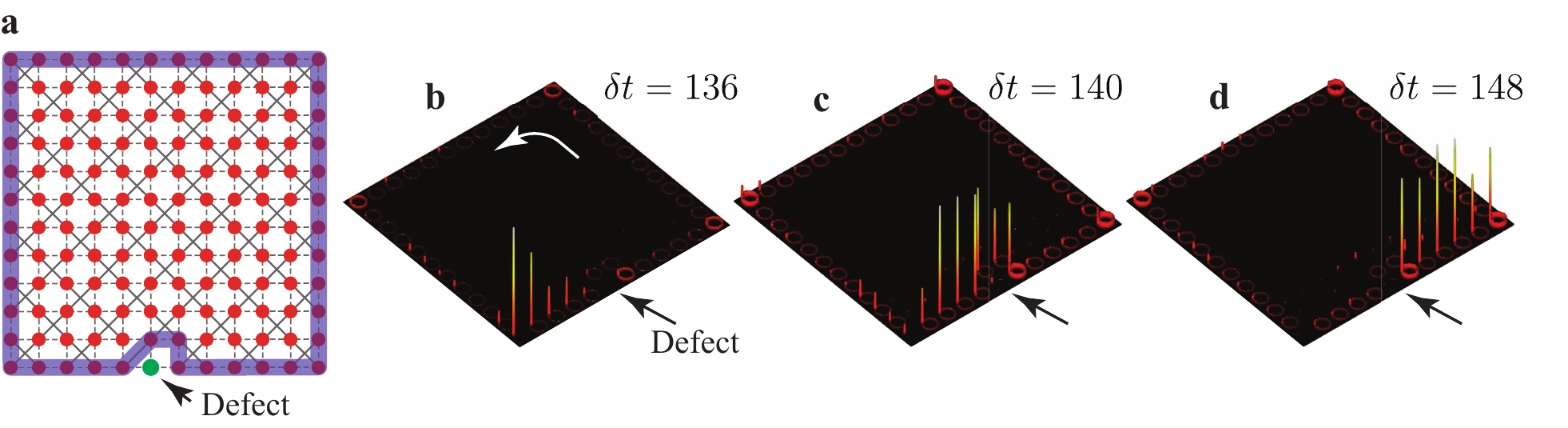}
 \caption {
 \textbf{a} Schematic of the lattice with a deliberately located defect on the boundary.
 \textbf{b-d} Robust propagation of single super-solitons around the defect, at different slow times, without any scattering into the bulk. Also see Movie M3.
 }
 \label{fig:4}
\end{figure*}

In the linear regime, the edge states have been demonstrated to be topologically protected against defects in the lattice \cite{Hafezi2013, Mittal2014, Mittal2019}. To investigate whether the edge states preserve their robustness in the nonlinear regime as well, we explore the propagation of super-solitons in the presence of a deliberately located point defect in the lattice. Specifically, we detune one of the site-ring resonators on the edge of the lattice by $20 J$ (Fig.\ref{fig:4}\textbf{a}) such that it is effectively decoupled from the rest of the lattice. Figures \ref{fig:4}\textbf{b-d} show slow-time evolution of the observed single super-soliton in this lattice (also see Movie M3). We see that the super-soliton simply routes around the defect as it circulates along the boundary of the lattice. We do not observe any light pulses that are reflected back from the defect or scattered into the bulk of the lattice. This observation clearly shows that our super-solitons are indeed topologically robust against defects in the lattice. \\

\noindent
\textbf{Discussion and Outlook}

\begin{figure}
 \centering
 \includegraphics[width=0.48\textwidth]{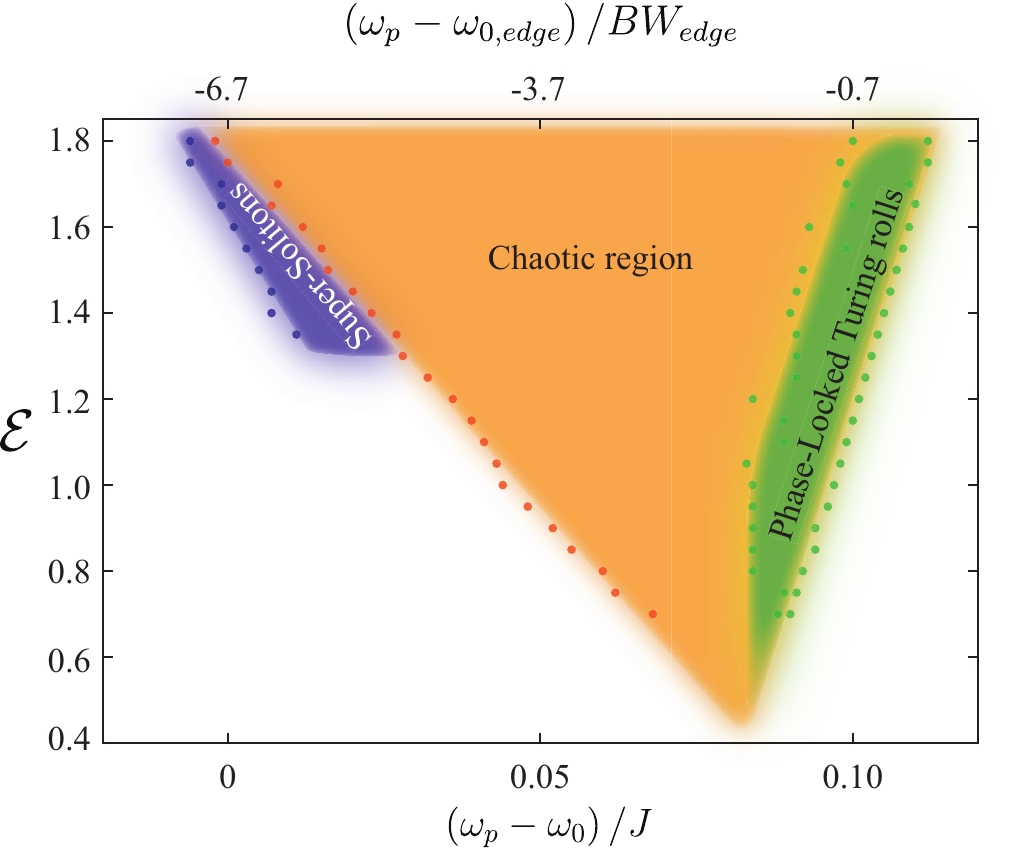}
 \caption {Qualitative phase diagram in the pumped edge mode, showing regions of Turing rolls, chaos, and super-solitons. The dots indicate numerical simulation results.
 }
 \label{fig:5}
\end{figure}

The emergence of coherent temporal features such as Turing rolls and super-solitons, and the characteristics of generated frequency comb in our topological super-ring resonator closely resembles that of a single ring resonator frequency comb in the anomalous dispersion regime. This allows us to qualitatively depict the phase diagram of our topological frequency comb as shown in Fig.\ref{fig:5}. We perform numerical simulations at different pump powers $\left(\E \right)$, and using spatio-temporal (fast-time) intensity distributions integrated over the edge rings (similar to Figs.\ref{fig:2}\textbf{d} and Fig.\ref{fig:3}\textbf{a}), we locate the regions of pump frequency detuning that lead to Turing rolls, chaos, and super-solitons. On this phase diagram, we have also indicated (upper $X-$axis) the pump frequency detuning from the respective cold-cavity (linear) edge mode resonance ($\omega_{o,edge}$ and normalized it by its bandwidth ($BW_{edge}$). Similar to the case of a single ring resonator (see Fig.\ref{fig:1}\textbf{d} and Supplementary Fig.S2), we observe Turing rolls at low pump powers and for pump frequencies near the cold-cavity edge resonance ($\omega_{0,edge}$). At pump frequencies further away from the cold-cavity edge resonance, we mostly observe a chaotic region. The soliton region appears at input pump power $\E \approx 1.3$, and at pump frequencies that are far red-detuned from the cold-cavity edge resonance. Furthermore, we observe that the soliton region narrows down, and completely disappears at higher pump powers, probably leading to another chaotic regime. We emphasize that this qualitative phase diagram was estimated for a given value of dispersion, in one of the edge mode resonances near the center of the edge band.

From an experimental perspective, our topological coupled ring resonator design has already been implemented on silicon-on-insulator platform, and has been used to generate spectrally engineered energy-time entangled photon pairs via spontaneous-four wave mixing \cite{Mittal2019, Orre2020, Mittal2018}. Some of the parameters used in our simulation, such as the coupling rates $J$ and $\kex$, can be tuned easily, to say, 5 GHz, and 250 MHz, respectively. The rest of the parameters, such the required loss rate $\kin = 25$ MHz (intrinsic quality factor of $\approx 8 \times 10^{6}$ at telecom wavelengths), dispersion $D_{2} = 50$ MHz, and pump power of the order of few Watts, can also be easily achieved on ultra low-loss silicon-nitride platform \cite{Kippenberg2018, Pasquazi2018}. Therefore, a major consideration is the intrinsic fabrication-induced disorder in the ring resonance frequencies, coupling strengths, and hopping phases. While here we have shown the robustness of topological Kerr super-solitons only against discrete defects in the lattice, our topological ring resonator design has been experimentally demonstrated to be robust against such random disorders that exist in practical systems \cite{Mittal2014, Mittal2018}. In fact, the edge states in our design are robust as long as the disorder strength is less than the width of the topological bandgap $\left( \sim 2 J\right)$ \cite{Mittal2014, Mittal2018}. Therefore, we believe our topological super-soliton frequency comb can be implemented using existing integrated photonic platforms and state of the art nanofabrication technology.

While we shown the presence of many features that are analogous to a single ring resonator frequency comb, we have only analyzed a small subset of the parameters that control our topological frequency comb. Therefore, one can expect appearance of many other known and unknown phases that could emerge from the interaction of edge and bulk modes. It would be intriguing, for example, to explore breathing Turing rolls and super-solitons, dark super-solitons and platicons in the normal dispersion region \cite{Chembo2013, Pasquazi2018}. In the limit of weak pump powers, our results could pave the way for generation of quantum-optical frequency combs and photonic cluster states entangled in higher dimensions using frequency-time multiplexing \cite{Kues2017, Reimer2019}. Our system could be translated to other frequency regimes of the electromagnetic spectrum, for example, to the microwave domain using circuit QED platform to implement topological arrays of coupled resonators \cite{Carusotto2020}. One could also explore other topological lattice models to engineer the band structure, and thereby, the dispersion of edge and bulk states. In fact, one could go beyond Euclidean geometries, and explore the hierarchy of solitons in non-Euclidean curved space, for example, the hyperbolic lattices \cite{Kollar2019}.\\

\noindent
\textbf{Acknowledgements}
This research was supported by the Air Force Office of Scientific Research Multi University Research Initiative (AFOSR-MURI grant FA9550-16-1-0323), Office of Naval Research Multi University Research Initiative (ONR-MURI grant N00014-20-1-2325), U. S. Army Research Laboratory grant W911NF1920181, and NSF grant PHY1820938.

\bibliographystyle{NatureMag}
\bibliography{FC_Biblio,Topo_Biblio}

\end{document}



\title{Supplementary Information: Topological Frequency Combs and Nested Temporal Solitons}
\author{Sunil Mittal}
\email[Email: ]{mittals@umd.edu}
\affiliation{Joint Quantum Institute, NIST/University of Maryland, College Park, MD 20742, USA}
\affiliation{Department of Electrical and Computer Engineering, and IREAP, University of Maryland, College Park, MD 20742, USA}

\author{Gregory Moille}
\affiliation{Microsystems and Nanotechnology Division, Physical Measurement Laboratory, National Institute of Standards and Technology, Gaithersburg, MD 20899, USA}
\affiliation{Joint Quantum Institute, NIST/University of Maryland, College Park, MD 20742, USA}

\author{Kartik Srinivasan}
\affiliation{Microsystems and Nanotechnology Division, Physical Measurement Laboratory, National Institute of Standards and Technology, Gaithersburg, MD 20899, USA}
\affiliation{Joint Quantum Institute, NIST/University of Maryland, College Park, MD 20742, USA}

\author{Yanne K. Chembo}
\affiliation{Department of Electrical and Computer Engineering, and IREAP, University of Maryland, College Park, MD 20742, USA}

\author{Mohammad Hafezi}
\affiliation{Joint Quantum Institute, NIST/University of Maryland, College Park, MD 20742, USA}
\affiliation{Department of Electrical and Computer Engineering, and IREAP, University of Maryland, College Park, MD 20742, USA}
\affiliation{Department of Physics, University of Maryland, College Park, MD 20742, USA}

\maketitle

\section{Spectrum of the linear topological ring resonator lattice}

\begin{figure*} [ht]
 \centering
 \includegraphics[width=0.98\textwidth]{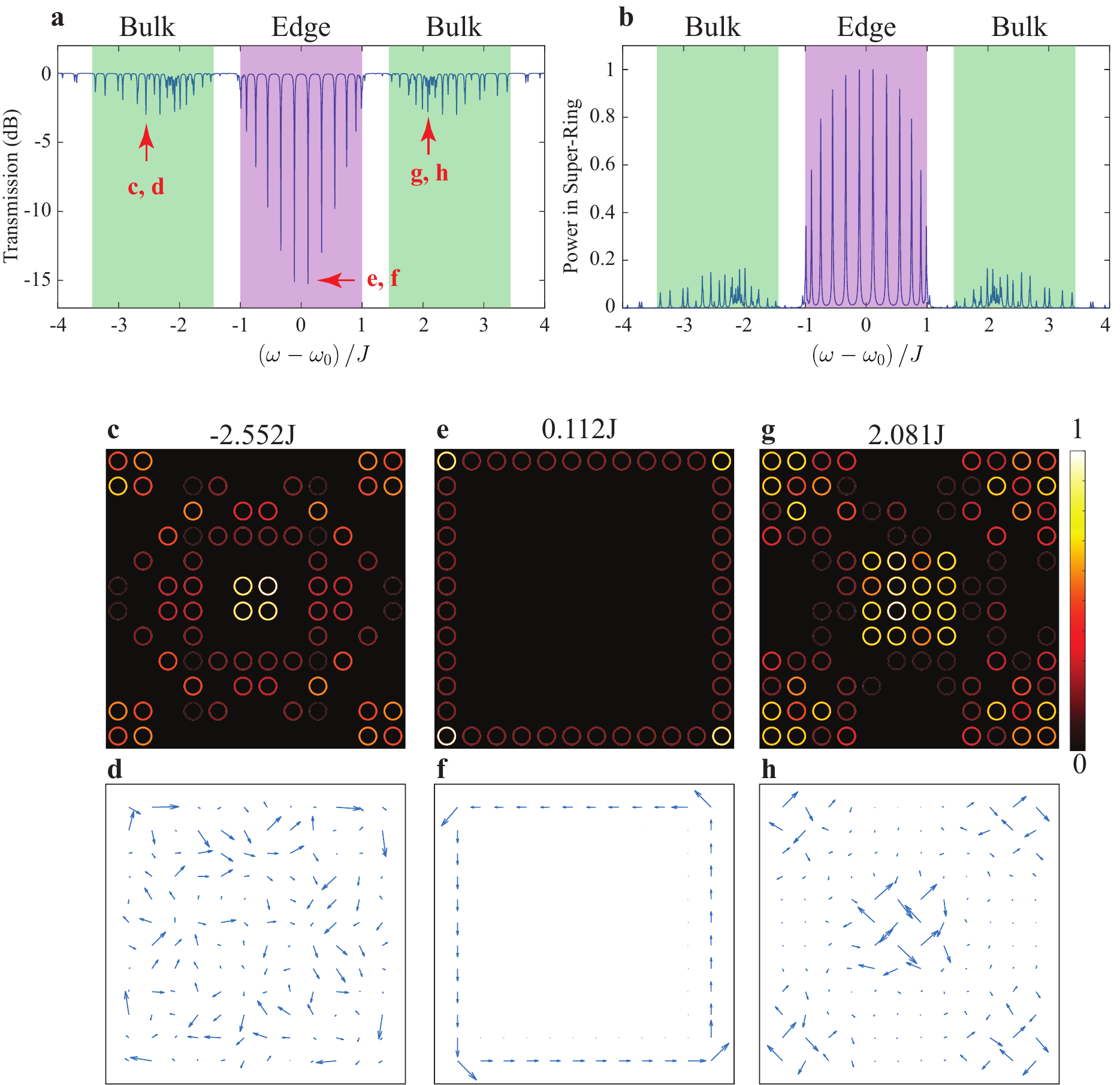}
 \caption {
 \textbf{a} Linear transmission spectrum of the topological ring resonator lattice shown in Fig.1\textbf{d} of the main text.
 \textbf{b} Total power (normalized) in the super-ring resonator formed by rings on the edge of the lattice. Here $\omega$ is the frequency of the input light and $\omega_{0}$ is the longitudinal mode resonance frequency of the individual ring resonators. The edge and the bulk bands are highlighted in purple and green, respectively.
 Spatial intensity distribution, and the direction of flow of photons in the lattice when the input light frequency is, \textbf{c,d} in the left bulk band, at $\omega - \omega_{0} = -2.552 J$.
 \textbf{e,f} in the edge band, at $\omega - \omega_{0} = 0.112 J$, and \textbf{g,h} in the right bulk band, at $\omega - \omega_{0} = 2.081 J$. In the bulk bands, photons occupy most of the bulk of the lattice and the flow of photons in the lattice is random indicating that the photons do not have a well-defined momentum. In the edge band, the intensity distribution of photons is confined to the edge of the lattice and the photons circulate around the lattice in a counter-clockwise direction, implementing a super-ring resonator. The peaks in the edge band of Fig.\ref{fig:S1}\textbf{a,b} are the longitudinal modes of this super-ring resonator. The plots here show only the site ring resonators.
}
 \label{fig:S1}
\end{figure*}

In this section we discuss the properties of our topological ring-resonator lattice (shown in Fig.1\textbf{d} of the main text) in the linear regime. Fig.\ref{fig:S1}\textbf{a,b} show the transmission spectrum and total power (normalized) in the super-ring resonator formed by edge states, as a function of the input light frequency detuning $\left(\omega-\omega_{0}\right)$. The field at the output of the lattice, $\mathcal{E}_{out}$, is calculated using the input-output formalism as
\begin{equation}
\mathcal{E}_{out} = \mathcal{E}_{in} + i \sqrt{2 \kex} a_{IO},
\end{equation}
where $\kex$ is the coupling rate of the input-output waveguide to the input-output ring, $\mathcal{E}_{in}$ is the input field, and $\left| a_{IO} \right|^{2}$ is total energy in the input-output ring. The total power (or energy) in the super-ring resonator is calculated as $\sum_{m \in edge} \left| a_{m} \right|^{2}$. This transmission spectrum repeats after every FSR $\OR$ of the individual ring resonators.

The frequency spectrum of our topological devices shows one edge band sandwiched between a pair of bulk bands \cite{Leykam2018, Mittal2019}. When the input light frequency is in the bulk bands, its spatial intensity distribution in the lattice is spread through the bulk of the lattice (Figs.\ref{fig:S1}\textbf{b,f}). Furthermore, the photons do not have a well defined momentum or direction of propagation in the lattice (Figs.\ref{fig:S1}\textbf{c,g}). In contrast, when the input light frequency corresponds to the edge band, its spatial intensity distribution is confined only to the edge of the lattice. Furthermore, photons in the edge band circulate around the lattice in counter-clockwise direction, and thereby, realize a travelling-wave super-ring resonator. The higher intensity observed at the corners of the lattice is because these sharp corners act as defects in the lattice. Nevertheless, these defects can not scatter edge states into the bulk of the lattice because the edge states are topologically protected.

We note that our system is time-reversal symmetric, and supports a pseudo-spin degree of freedom that corresponds to the circulation direction of photons in the ring resonators \cite{Hafezi2011, Hafezi2013, Leykam2018, Mittal2019}. We can excite a single pseudo-spin in this system by appropriately  choosing the input to the lattice. Furthermore, in the absence of backscattering, the two pseudo-spins are decoupled and our system simulates the quantum spin-Hall effect. The edge states corresponding to the two pseudo-spins circulate around the lattice in opposite directions. The plots shown here are for the case when photons circulate clockwise in the site-ring resonators.

\section{Optical frequency comb and dissipative Kerr-solitons in a single ring resonator}

In this section we discuss the generation of an optical frequency comb, and temporal features such as Turing rolls and Kerr solitons in a single ring resonator, in an all-pass filter geometry (Fig.1\textbf{a} of the main text). This discuss serves to highlight the similarities between a single-ring resonator frequency comb and our topological frequency comb that uses super-ring resonator formed by the edge states.

\begin{figure*}[h]
 \centering
 \includegraphics[width=0.98\textwidth]{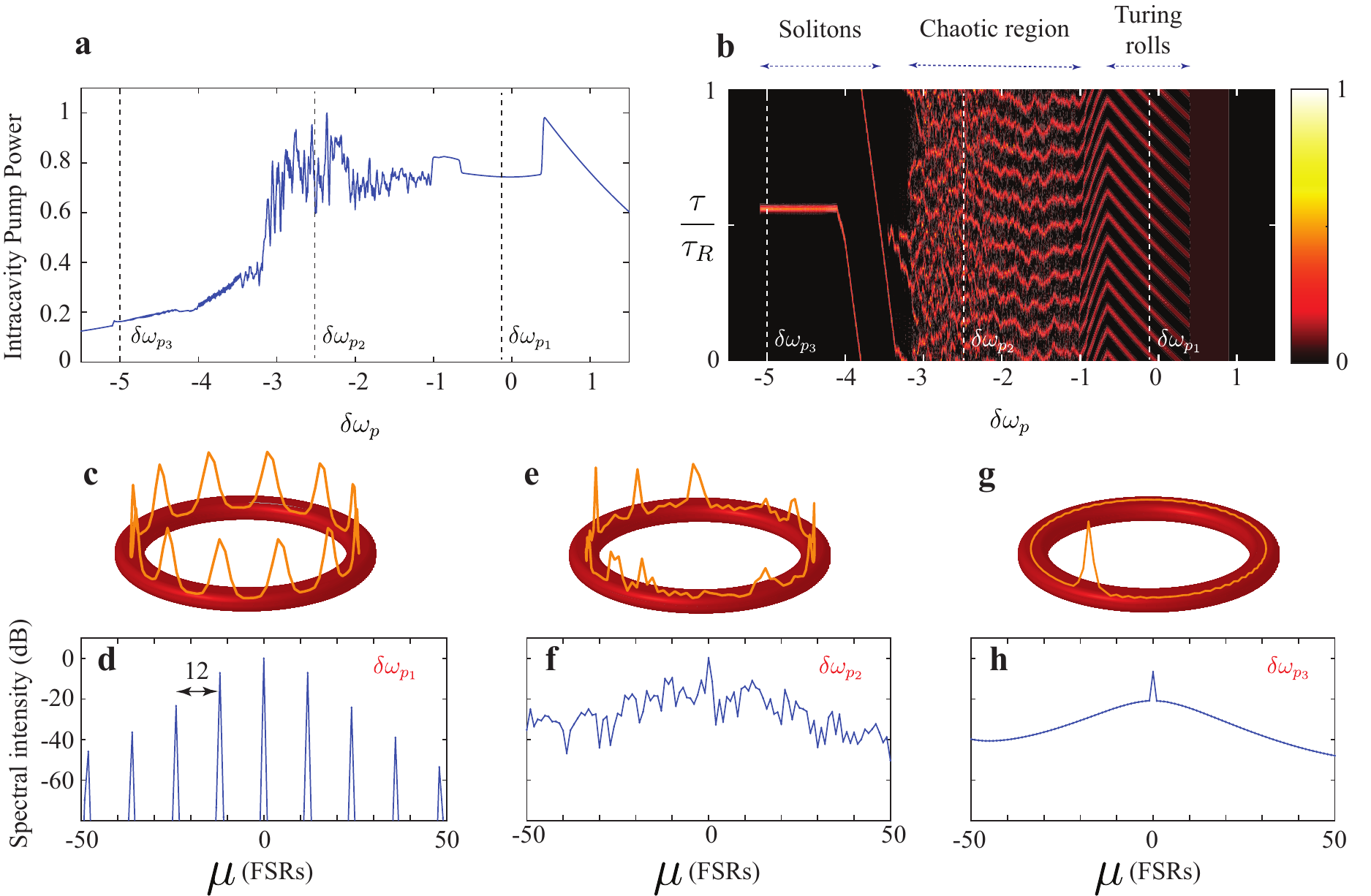}
 \caption {
 \textbf{a} Pump power in the ring resonator as a function of the input pump frequency detuning $\delta\omega_{p}$, normalized by the bandwidth $\left(\kex + \kin \right)$ of the all-pass filter. $\delta\omega_{p} = 0$ represents the cold-cavity resonance.
 \textbf{b} Spatio-temporal intensity distribution in the ring as a function of the fast-time $\tau / \TR$ and input pump frequency.
 Spatial intensity distribution and frequency comb spectrum, \textbf{c,d} in the region of Turing rolls, at $\delta\omega_{p_{1}}$; \textbf{e,f} in the chaotic region, at $\delta\omega_{p_{2}}$; and \textbf{g,h} in the region of solitons, at $\delta\omega_{p_{3}}$.
}
 \label{fig:S2}
\end{figure*}

Fig.\ref{fig:S2}\textbf{a} shows the the normalized pump power in the ring as a function of the normalized input pump frequency detuning $\delta \omega_{p} = \frac{\left(\omega_{p} - \omega_{0}\right)} {\left(\kex + \kin \right)}$. Here $\omega_{0}$ is the cold cavity resonance in the pump FSR, and $\kex$ and $\kin$ are the coupling and loss rates respectively. $\left( \kex + \kin \right)$ is the bandwidth of the all-pass filter in the linear regime. This spectrum was calculated using the Lugiato-Lefever formalism \cite{Chembo2013, Kippenberg2018, Pasquazi2018}. Fig.\ref{fig:S2}\textbf{b} shows the corresponding temporal intensity distribution in the ring as a function of fast-time $\tau / \TR$. Because the Lugiato-Lefever formalism uses a frame that is rotating at $\OR = \frac{1}{\tau_{R}}$, the round-trip time of the ring, the temporal intensity distribution is synonymous with the spatial intensity distribution in the ring.

We analyze three input pump frequencies in three different regions, as indicated in Figs.\ref{fig:S2}\textbf{a,b}. At $\delta\omega_{p_{1}}$, we observe the formation of Turing rolls in the rings (Fig.\ref{fig:S2}\textbf{c}). The corresponding frequency comb spectrum, in Fig.\ref{fig:S2}\textbf{d}, shows the presence of only a discrete set of longitudinal modes, separated by 12 FSRs. This frequency spacing (12) also corresponds to the number of Turing rolls in the ring resonator. Furthermore, the region of Turing rolls appears near $\delta\omega_{p} = 0 $, and the pump spectrum in this region shows a smooth behaviour. These plots can be compared with Figs.2\textbf{c,d,e,g} of the main text which show similar Turing rolls that are phase-locked across each ring on the edge of the lattice, a similar frequency comb spectrum, and similar smoothly varying pump spectrum in the region of Turing rolls.

At $\delta\omega_{p_{2}}$, we observe a chaotic region where the pump spectrum varies rapidly as a function of pump frequency. In this region, the spatial intensity profile in the ring is chaotic and does not show any distinct features (Figs.\ref{fig:S2}\textbf{e}). Furthermore, this is not a stationary solution meaning that the profile changes with slow-time evolution. The corresponding frequency comb, in Fig. \ref{fig:S2}\textbf{f}, also shows random variations, but in contrast to Turing rolls, all the longitudinal modes are oscillating (Figs.\ref{fig:S2}\textbf{f}). These plots can be compared with those of Fig.2\textbf{f,h,l} of the main text where we see a random spatial intensity distribution which varies from ring to ring, and non-stationary solutions at the output.

At $\delta\omega_{p_{3}}$, the pump spectrum shows step-like features. In this region we observe a single soliton in the ring resonator (Figs.\ref{fig:S2}\textbf{g}), with a frequency comb spectrum that is coherent and smooth across FSRs (Figs.\ref{fig:S2}\textbf{h}). This is a stationary solution meaning that the intensity profile in the ring does not change with slow-time evolution. These plots can be compared with plots in Fig.3\textbf{a-d} of the main text where we observe similar step-like features in the pump spectrum, and a super-soliton which is further made of phase-locked solitons in the individual ring resonators.

\section{Number of Turing Rolls as a function of Dispersion}

\begin{figure*}
 \centering
 \includegraphics[width=0.98\textwidth]{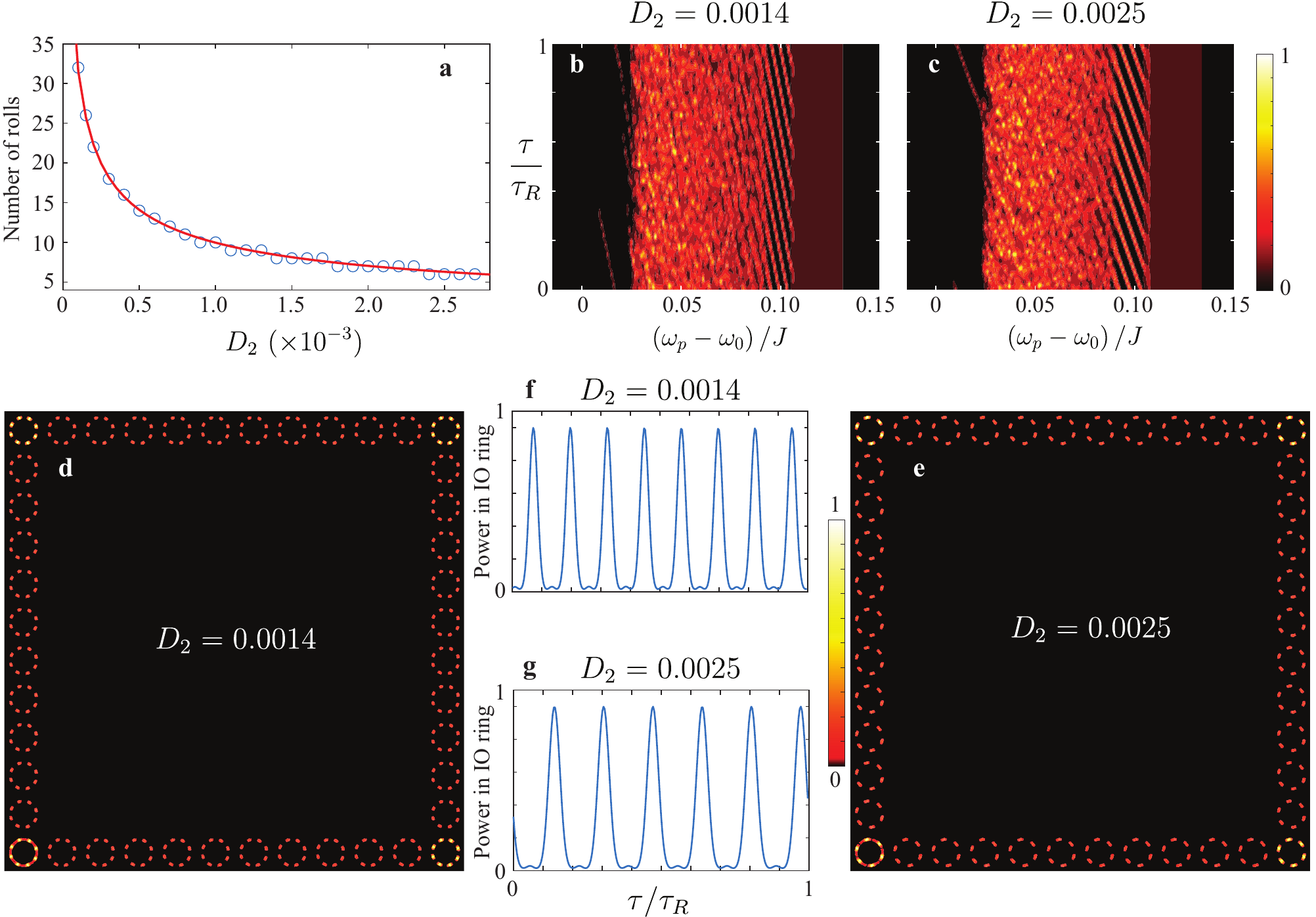}
 \caption {
 \textbf{a} Number of Turing rolls in the topological frequency comb as a function of the dispersion parameter $D_{2}$. Open circles are numerically simulated values and the solid-line is a $1/\sqrt{D_{2}}$ fit to the data.
\textbf{b,c} Spatial intensity distributions integrated over the super-ring resonator, for $D_{2} = 0.0014$ and $D_{2} = 0.0025$, respectively. Similar to Fig. 2\textbf{d} and Fig. 3\textbf{a} of the main text, we observe distinct regions of Turing rolls and also solitons as we tune the pump frequency.
\textbf{d,e} Spatial intensity distributions in the lattice for input pump frequencies in the region of Turing rolls. The Turing rolls are phase locked throughout the edge of the lattice.
\textbf{f,g} Turing rolls within a ring resonator on the edge, for the two dispersion values.
}
 \label{fig:S3}
\end{figure*}

For a single ring resonator frequency comb in the anomalous dispersion regime, the number of Turing rolls $N_{\text{Rolls}}$ decreases with increasing dispersion as $1/\sqrt{D_{2}}$, where $D_{2}$ is the second-order dispersion coefficient \cite{Chembo2010, Chembo2013}. We observe identical behaviour in our topological frequency comb. In Fig.\ref{fig:S3}\textbf{a} we plot the number of Turing rolls in the topological frequency comb, as we vary the dispersion parameter $D_{2}$. A fit (solid-line) to the numerically simulated data (open circles), shows that indeed $N_{\text{Rolls}} \propto 1 / \sqrt{D_{2}}$.

To show self-organization of the ring resonators on the edge of the lattice across a range of dispersion values, in Figs.\ref{fig:S3}\textbf{b,c} we plot the integrated spatio-temporal intensity distributions (integrated over the super-ring resonator) as a function of the input pump frequency, for $D_{2} = 0.0014$ and $D_{2} = 0.0025$, respectively. Figs.\ref{fig:S3}\textbf{d,e} show complete spatial intensity distributions in the lattice when the input pump frequency is in the region of Turing rolls $\left(\w_{p} - \w_{0} = 0.1 J \right)$. Similar to Fig.2 of the main text, we observe that the Turing rolls in all the ring resonators on the edge are locked to the same spatio-temporal phase. Similarly, we also observe the presence of super-solitons. Note that the dispersion values used here are 5-10 times of that used in Fig.2 and Fig.3 of the main text.

\section{Phase coherence of the super-soliton frequency comb}

\begin{figure*}
 \centering
 \includegraphics[width=0.98\textwidth]{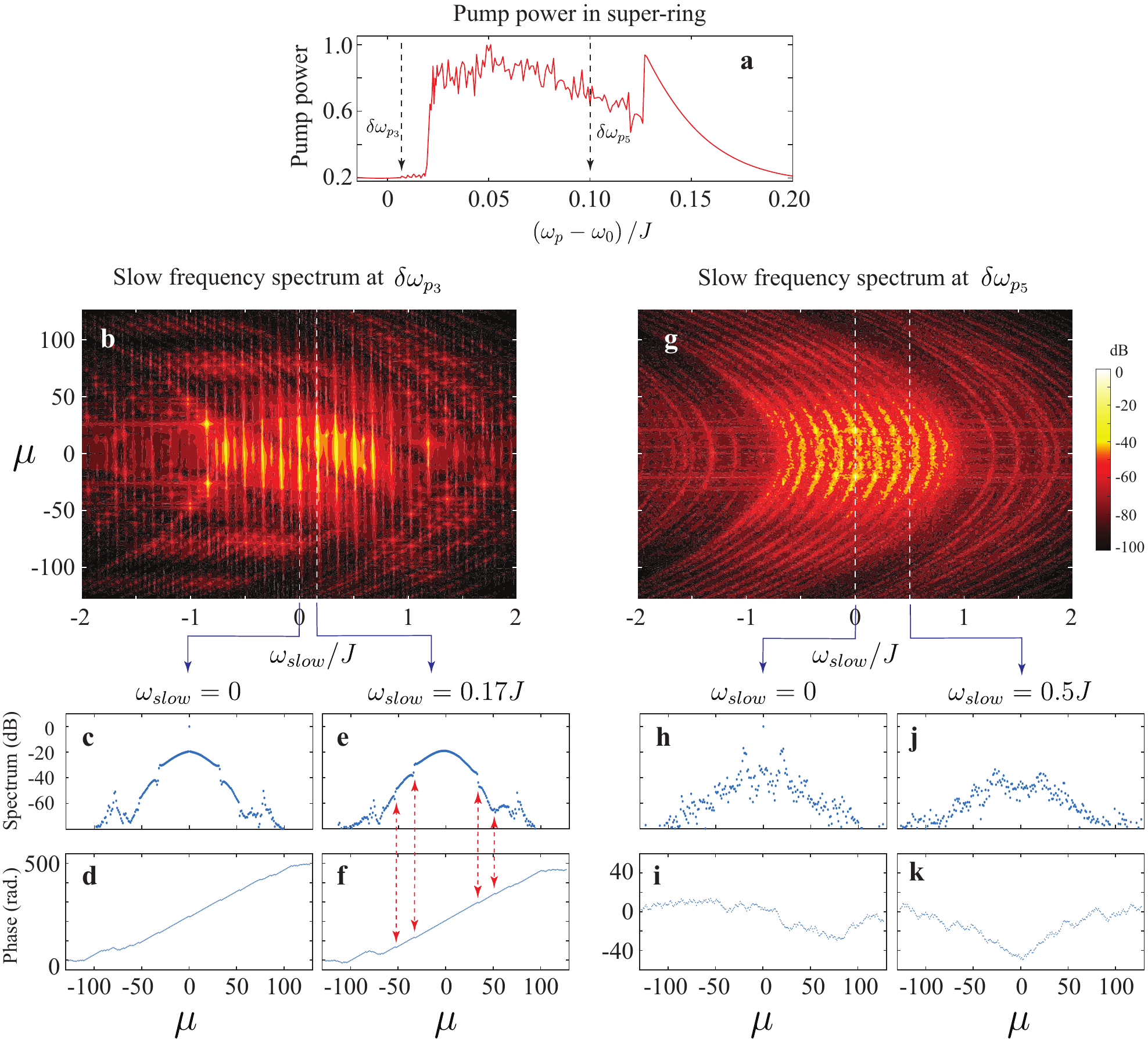}
 \caption {
 \textbf{a} Total pump power in the super-ring resonator as a function of the pump frequency detuning $\delta\omega_{p}$ (same as Fig. 3\textbf{b} of the main text, repeated here for convenience).
 \textbf{b} Slow-frequency spectrum in the regime of super-solitons (same as Fig. 3\textbf{k} of the main text).
 \textbf{c-f} Power spectrum and unwrapped phase of the comb along fast-frequency ($\mu$), at two edge state resonances with slow-frequencies as indicated on the figure. The power spectrum is smooth, except for kinks where different edge modes interfere because of dispersion. The phase varies linearly, with phase jumps exactly at locations of kinks in the power spectrum (shown by dashed red lines).
\textbf{g-k} Corresponding results for pump frequency in the chaotic regime. The comb spectrum is not smooth, and the phase varies randomly along the fast axis.
}
 \label{fig:S4}
\end{figure*}

In this section we show the phase coherence of our topological frequency comb in the regime of super-solitons. Fig.\ref{fig:S3}\textbf{a} show the integrated pump power in the super-ring resonator at $\mathcal{E} = 1.56$ (same as Fig.3\textbf{a} of the main text, repeated here for convenience). Figures \ref{fig:S3}\textbf{b,c} show the output frequency spectrum (both slow and fast frequencies) of the comb at pump frequencies $\delta\w_{p_{3}} = 0.007 J$ in the super-soliton region, and at $\delta\w_{p_{5}} = 0.10 J$ in the chaotic region. As we discussed in the main text, the comb spectrum in the super-soliton regime (Fig.\ref{fig:S3}\textbf{b}) shows oscillations of multiple equidistant edge modes with in a single FSR $\OR$ of the individual ring resonators. Furthermore, we observe cancellation of linear dispersion by nonlinearity-induced dispersion across multiple single ring FSRs (indexed by $\mu$, the fast frequency). In contrast, the comb spectrum in the chaotic region (Fig.\ref{fig:S3}\textbf{c}) also shows oscillation of multiple modes, but no cancellation of dispersion.

In a single ring-resonator solitonic frequency comb, cancellation of dispersion leads to phase coherence such that the output power varies smoothly across the comb lines \cite{Kippenberg2018, Pasquazi2018, Chembo2010, Chembo2013}. To show that our topological frequency comb is also phase coherent in the regime of super-solitons, in Figs.\ref{fig:S4}\textbf{d-g} we plot the spectral power and the phase of two oscillating edge modes resonances (at $\omega_{slow} = 0, 0.17 J$) along the fast-frequency axis (indexed by integer $\mu$, the FSRs of the individual ring resonators). The spectral power indeed varies smoothly across the comb lines in a region $-31 < \mu < 31$. However, near $\mu \approx \pm 32 \text{and} \pm 50$, and more frequently at higher $\mu$, we observe jumps in the comb spectrum. From Fig.\ref{fig:S4}\textbf{b}, we see that these jumps in the spectrum appears when two edge modes interfere (the start-like patterns in Fig.\ref{fig:S4}\textbf{b}).

The corresponding phase (Fig.\ref{fig:S4}\textbf{e,g}) of the comb lines in the super-soliton region also varies linearly in the regime $-31 < \mu < 31$ which shows that our comb is indeed phase-coherent. Nevertheless, we observe phase jumps exactly at locations of jumps in the comb spectrum (dashed red-lines in Fig.\ref{fig:S4}\textbf{e,g}). Within regions between two jumps, the phase variation is again linear. In contrast, the frequency comb in the chaotic region (Fig.\ref{fig:S4}\textbf{g-j}) exhibits randomly varying power and phase across the comb lines, indicating that the comb is not coherent. Note that here we have plotted the unwrapped phase to explicitly show its linear profile.

\section{Pumping another edge state resonance}

\begin{figure*}
 \centering
 \includegraphics[width=0.98\textwidth]{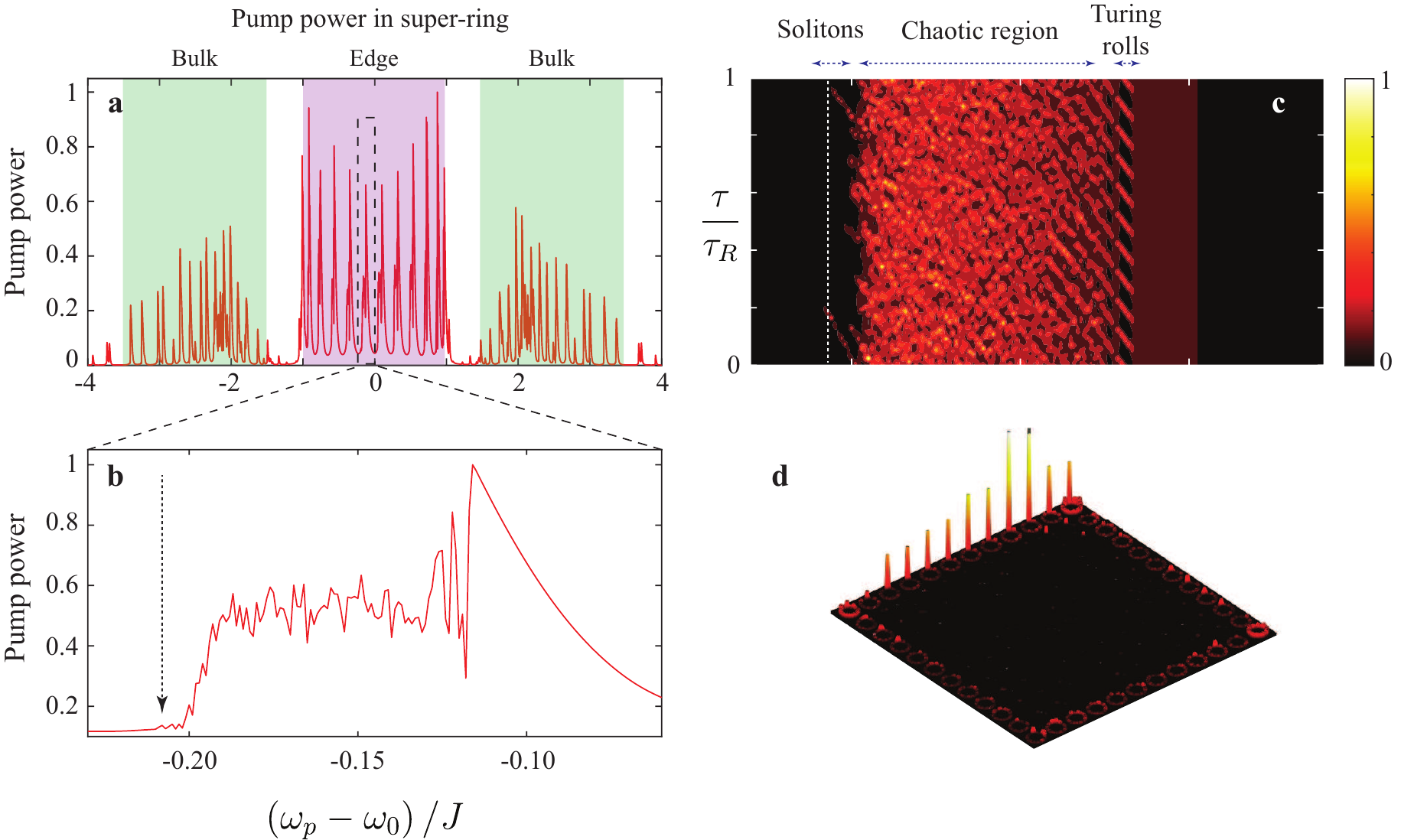}
 \caption {
 \textbf{a} Total pump power in the super-ring resonator as a function of input pump frequency.
 \textbf{b} A zoom-in of \textbf{a} in one of the edge state resonances.
 \textbf{c} Spatio-temporal intensity distribution in ring resonators, integrated over all the rings in the super-ring resonator. Similar to Fig.2\textbf{d} and Fig.3\textbf{a} of the main text, the intensity distribution shows a region of phase-locked Turing rolls, a chaotic region, and a region of dissipative Kerr super-solitons.
 \textbf{d} Spatial intensity profile in the lattice showing a super-soliton. The input pump frequency is indicated by dashed lines in \textbf{b} and \textbf{c}.
}
 \label{fig:S5}
\end{figure*}

In the main text, we presented results for pumping one of the edge state resonances in the middle of the edge band $\left(\approx \left[0, 0.2\right] J \right)$. Fig.\ref{fig:S5} shows simulation results for pumping another edge state resonance, in the range $\approx \left[-0.25, -0.05\right] J$ (indicated by dashed rectangle in Fig.\ref{fig:S5}\textbf{a}). Figure \ref{fig:S5}\textbf{b} shows the total pump power in the super-ring resonator, and Fig.\ref{fig:S5}\textbf{c} shows the spatio-temporal (fast time) intensity distribution in the ring resonators, integrated over all the rings in the super-ring resonator. These plots are very similar to those shown in Fig.2 and Fig.3 of the main text. As before, the spatial intensity distribution shows a region of Turing rolls that are phase-locked through out the rings on the edge of the lattice, and a region of super-solitons. Fig.\ref{fig:S5}\textbf{d} shows an example of the super-solitons at pump frequency indicated by dashed lines in Figs.\ref{fig:S5}\textbf{b,c}. Here, we chose $D_{2} = 0.001$, and $\mathcal{E} = 1.47$.

\section{Dispersion of Edge States}

\begin{figure*}
 \centering
 \includegraphics[width=0.98\textwidth]{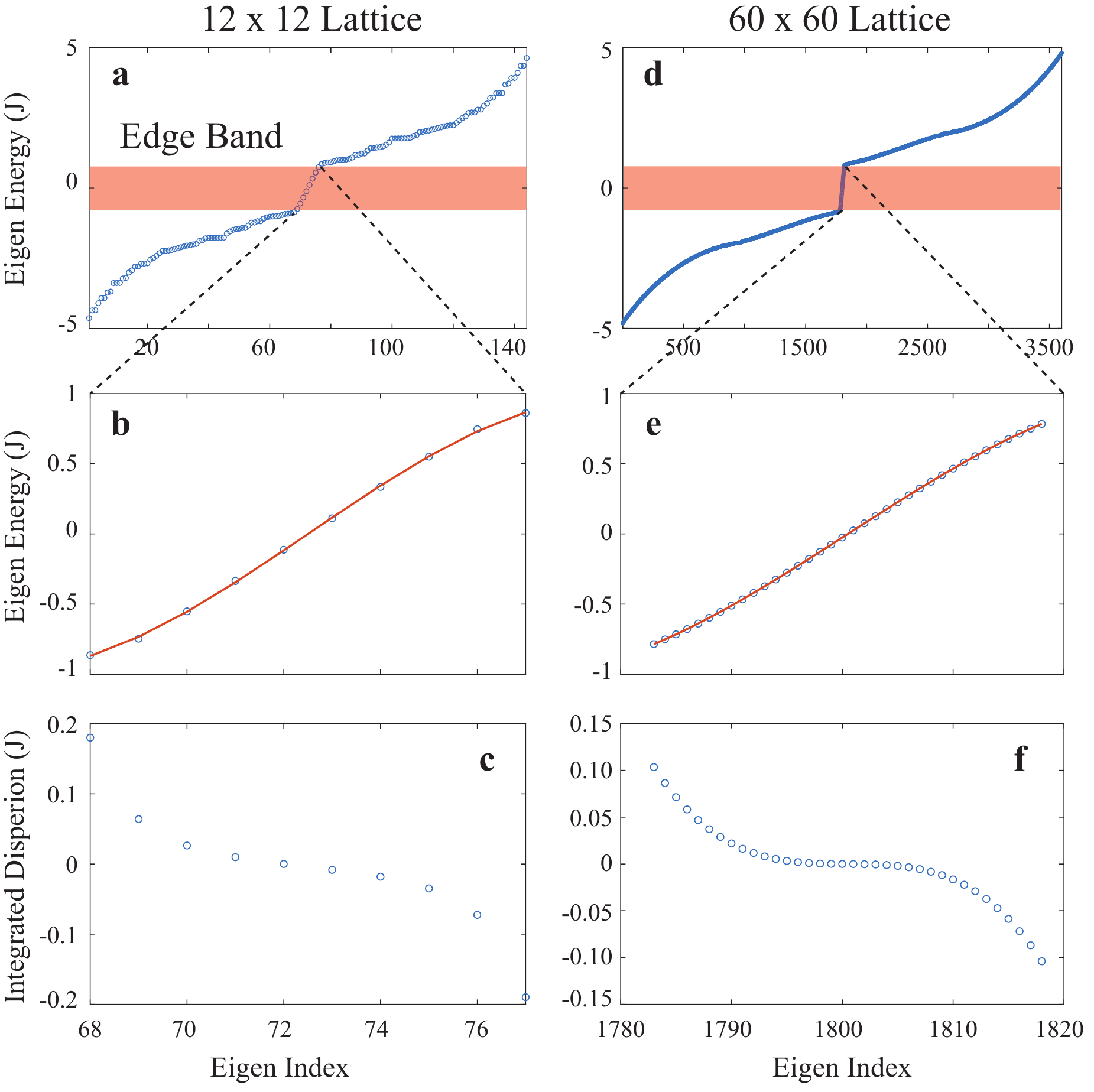}
 \caption {
 \textbf{a} Energy eigenvalues for a $12 \times 12$ lattice of (site) ring resonators in the linear regime. The edge band $\approx \left(-1, 1 \right)J$ is shaded in red.
 \textbf{b} A zoom-in of \textbf{a} in the edge band region. The solid-line is a third-order polynomial fit to the eigenvalues (circles).
 \textbf{c} Integrated dispersion of the edge state resonances, showing the predominance of third-order dispersion.
 \textbf{d-f} Corresponding results for a $60 \times 60$ lattice.
}
 \label{fig:S6}
\end{figure*}

In our numerical simulations, we include second-order anomalous dispersion $D_{2}$ for the ring resonance frequencies $\omega_{0,\mu}$. As we mentioned in the main text, the dispersion of the edge states resonances (longitudinal modes of the super-ring resonator) within a single FSR of the individual ring resonators is completely determined by the tight-binding Hamiltonian (Eq.(1) of the main text) that describes the anomalous quantum Hall effect. Figure \ref{fig:S6}\textbf{a} shows the energy eigenvalues for this Hamiltonian, calculated for a $12 \times 12$ lattice of ring (site) resonators. Figure \ref{fig:S6}\textbf{b} shows a zoom-in of the eigenvalues in the edge band ($\approx\left(-1, +1 \right) J$). While the edge states in the center of the edge band can be well described by a linear dispersion, as is usually accepted in the condensed-matter community, we see that the edge state resonances closer to the bulk bands do exhibit significant nonlinear dispersion.

We use a third-order polynomial to fit the edge state resonances (shown in Fig.\ref{fig:S6}\textbf{b} as solid-line) such that $\omega_{0,\nu} = \omega_{0,\nu_{0}} + \OSR \left(\nu - \nu_{0} \right) + \frac{D_{2}^{edge}}{2} \left(\nu - \nu_{0} \right)^2 + \frac{D_{3}^{edge}}{6} \left(\nu - \nu_{0} \right)^3$. Here $\omega_{0,\nu}$ are the edge state resonance frequencies indexed by $\nu$, $\nu_{0}$ is the resonance in the center of the edge band (here 72nd eigenvalue), $D_{2}^{edge}$ and $D_{3}^{edge}$ are the second- and third-order dispersion coefficients, respectively. For the $12 \times 12$ lattice, we find $\OSR = 0.2328 J$,  $D_{2}^{edge} = -0.0062 J$, and $D_{3}^{edge} = -0.0123 J$. Figure \ref{fig:S6}\textbf{c} shows integrated dispersion that is $\frac{D_{2}^{edge}}{2} \left(\nu - \nu_{0} \right)^2 + \frac{D_{3}^{edge}}{6} \left(\nu - \nu_{0} \right)^3$. Therefore, we find that the dispersion of the edge states is in fact dominated by the third-order dispersion $D_{2}^{edge}$. Furthermore, $D_{2}^{edge} / \OSR \simeq -26 \times 10^{-3}$, and $D_{3}^{edge} / \OSR \simeq -53 \times 10^{-3}$.

Figures \ref{fig:S6}\textbf{d-f} show corresponding results for a $60 \times 60$ lattice. In this case, we find $\OSR = 0.0508 J$,  $D_{2}^{edge} = -5.938 \times 10^{-5} J$, and $D_{3}^{edge} = -11.598 \times 10^{-5} J$. Note that the ratios $D_{2}^{edge} / \OSR \simeq -1.2 \times 10^{-3}$, and $D_{3}^{edge} / \OSR \simeq -2.3 \times 10^{-3}$. These are an order of magnitude smaller compared to that of the $12 \times 12$ lattice, suggesting that the effective nonlinear dispersion of the edge states decreases with increasing lattice size. This is because edge state resonances in the center of the edge band exhibit linear dispersion, and the nonlinear dispersion originates only from regions closer to the bulk bands. With increasing lattice size, the relative contribution of the regions closer to the bulk bands decreases. Note that the width of the edge band is $\simeq 2J$, irrespective of the size of the lattice.

\bibliographystyle{NatureMag}
\bibliography{FC_Biblio,Topo_Biblio}